

\newdimen\tempdim                 
\newdimen\othick   \othick=.4pt   
\newdimen\ithick   \ithick=.4pt   
\newdimen\spacing  \spacing=9pt   
\newdimen\abovehr  \abovehr=6pt   
\newdimen\belowhr  \belowhr=8pt   
\newdimen\nexttovr \nexttovr=8pt  

\def\rr{\hfil\down{\abovehr}&\omit\vrsp\vrule width\othick\cr
     \noalign{\hrule height\ithick}\up{\belowhr}&}
\def\up#1{\tempdim=#1\advance\tempdim by1ex
     \vrule height\tempdim width0pt depth0pt}
\def\down#1{\vrule height0pt depth#1 width0pt}
\def\large#1#2{\setbox0=\vtop{\hsize#1 \lineskiplimit=0pt \lineskip=1pt
     \baselineskip\spacing \advance\baselineskip by 3pt \noindent
     #2}\tempdim=\dp0\advance\tempdim by\abovehr\box0\down{\tempdim}}
\def\vrsp{\hskip\nexttovr\relax}
\def\toprule#1{\def\startrule{\hrule height#1\relax}} 
\toprule{\othick}                      
\def\nstrut{\vrule height\spacing depth3.5pt width0pt}
\def\preamble#1{\def\startup{#1}}      
\preamble{&##}                         
{\catcode`\!=\active
 \gdef!{\hfil\vrule width0pt\vrsp\vrule width\ithick\relax\vrsp&}}

\def\table #1{\vbox\bgroup \setbox0=\hbox{#1}
     \vbox\bgroup\offinterlineskip  \catcode`\!=\active
     \halign\bgroup##\vrule width\othick\vrsp&\span\startup\nstrut\cr
     \noalign{\medskip}
     \noalign{\startrule}\up{\belowhr}&}

\def\caption #1{\down{\abovehr}&\omit\vrsp\vrule width\othick\cr
     \noalign{\hrule height\othick}\egroup\egroup \setbox1=\lastbox
     \tempdim=\wd1 \hbox to\tempdim{\hfil \box0 \hfil} \box1 \smallskip
     \hbox to\tempdim{\advance\tempdim by-20pt\hfil\vbox{\hsize\tempdim
     \noindent #1}\hfil}\egroup}


\def\ut{{\tilde u}}
\def\vt{{\tilde v}}

\def\idu{ {\int_0^1 du\>}}
\def\idv{ {\int_0^1 dv\>}}
\def\idut{ {\int_0^1 d\ut\>}}
\def\idvt{ {\int_0^1 d\vt\>}}

\def\xt{{\tilde x}}
\def\b{{\beta}}
\def\a{{\alpha}}

PREPRINT $\#$ WISC-MILW-94-TH-10
\baselineskip 24pt
\vskip 0.2in
\centerline{\bf A CLOSED-FORM EXPRESSION FOR THE GRAVITATIONAL}
\centerline{\bf RADIATION RATE FROM COSMIC STRINGS}

\centerline{Bruce Allen}
\centerline{ Department of Physics, University of Wisconsin -- Milwaukee}
\centerline{P.O. Box 413, Milwaukee, Wisconsin 53201, U.S.A.}
\centerline{email: ballen@dirac.phys.uwm.edu}
\vskip0.2in
\centerline{Paul Casper}
\centerline{ Department of Physics, University of Wisconsin -- Milwaukee}
\centerline{P.O. Box 413, Milwaukee, Wisconsin 53201, U.S.A.}
\centerline{email: pcasper@dirac.phys.uwm.edu}
\vskip 0.2in
\noindent
\centerline{ ABSTRACT}
We present a new formula for the rate at which cosmic strings lose
energy into gravitational radiation, valid for all piecewise-linear
loops of infinitely-thin cosmic string .  At any time, such a loop is
composed of $N$ straight segments, each of which has constant
velocity.  Any cosmic string loop can be arbitrarily-well approximated
by a piecewise-linear loop with $N$ sufficiently large.  The formula is
a sum of $O(N^4)$ polynomial and log terms, and is exact when the
effects of gravitational back-reaction are neglected.  For a given
loop, the large number of terms makes evaluation ``by hand"
impractical, but a computer or symbolic manipulator yields accurate
results.  The formula is more accurate and convenient than previous
methods for finding the gravitational radiation rate, which require
numerical evaluation of a four-dimensional integral for each term in an
infinite sum.  It also avoids the need to estimate the contribution
from the tail of the infinite sum.  The formula has been tested against
all previously published radiation rates for different loop
configurations.  In the cases where discrepancies were found, they were
due to numerical errors in the published work.  We have isolated and
corrected the errors in these cases.  To assist future work in this area, a
small catalog of results for some simple loop shapes is provided.
\eject
\centerline{INTRODUCTION}
\vskip 0.05in
Cosmic strings are one-dimensional topological defects which appear in
some gauge theories of the fundamental interactions.  Strings would
appear at phase transitions where symmetries of the fundamental
interactions are spontaneously broken [1,2,3].  \ It is thought that
cosmic strings might have formed as the universe expanded and cooled
during the past.  They are remarkably simple objects, characterized by
a single parameter $\mu$, which is their mass-per-unit length.  For
strings of cosmological interest, the expected value of the
dimensionless parameter $G \mu/c^2$ is of order $10^{-6}$, where $G$ is
Newton's gravitational constant and $c$ is the speed of light.  The
strings of interest for this work are strings without ends - thus they
are always topologically in the form of circles, or possibly infinite
in length (in a spatially infinite universe).

The dynamics of a network of cosmic strings in an expanding universe
have been thoroughly studied [4,5,6].  To describe these dynamics, it
is useful to divide the strings, for the purposes of labeling, into two
categories:  the long string (length greater than the horizon length)
and the loops (all the rest).  Early work on cosmic strings established
that the energy-density of the long strings was a small constant
fraction (of order $G \mu/c^2$) of the energy-density of the
cosmological fluid.  In the literature, this is referred to as
``scaling" behavior.  The long string network maintains scaling
behavior by constantly ``chopping off" loops of cosmic string.  This
process takes place whenever long strings meet each other, make
contact, and ``intercommute".  Typically, after a loop is chopped off
it begins to oscillate due to its own tension, undergoing a process of
self-intersection (fragmentation) and eventually creating a family of
non-self-intersecting oscillating loops.   In the absence of
gravitational radiation, these loops would survive forever, oscillating
periodically, and would eventually come to dominate the energy-density
of the universe [1].  However, these loops gradually decay away due to
the emission of gravitational radiation [3].

The emission of gravitational radiation is thus of fundamental
importance to the topic of cosmic strings.   Indeed, the resulting
stochastic background of gravitational radiation left behind from the
families of small string loops provides the main cosmological
constraints on cosmic strings, through two observable effects [7
and references therein].  First, because gravitational
radiation contributes to the energy-density, it affects the expansion
rate of the universe.  The amount of gravitational radiation
must not be too great or it would interfere with the
highly-successful standard model of nucleosynthesis.  Secondly, the
amount of gravitational radiation must not be too great to interfere
with the extremely small timing residuals observed in the periods of a
number of carefully observed fast pulsars.  The work on these
cosmological constraints is reviewed and updated in [7].

During the past fifteen years a number of detailed calculations have
been carried out to determine the rate at which cosmic string loops
convert their energy into gravitational radiation.  The power radiated
by a given loop is $$P={E \over \Delta t}=\gamma G\mu^2 c,\eqno(1.1)$$
where $E$ is the energy radiated in gravitational waves in a single
oscillation of the loop, $\Delta t$ is the period of that oscillation
and $\gamma$ is a dimensionless constant that depends only on the shape
of the loop and its velocity at any fixed instant in time.  Thus the
problem is to determine the numerical value of $\gamma$ for a given
string loop.  Because loops are relativistic objects which have typical
velocities of order $c$, the simplest approximation formulae like the
quadrupole approximation are not of much use, although in some cases
they are reasonably accurate [8].  Vachaspati and Vilenkin [9] carried
out the first detailed calculation of $\gamma$ for a simple
generalization of the circular loop.  While some of the integrations
were carried out analytically, the final integration over directions
could only be done numerically.  The next work was a hybrid
analytic/numerical calculation by Burden [10], for a set of loops which
were a variation of the Vachaspati and Vilenkin family.  The first
entirely exact analytic calculations were done by Garfinkle and
Vachaspati [11], who considered a special family of ``kinky" string
trajectories.  These are the simplest piecewise linear loops for which
the exact formulae given in this paper may be applied directly.
Additional work by Durrer [8] repeated some of the earlier calculations
of the previous three groups and also investigated the accuracy of the
quadrupole approximation for determining $\gamma$.
The next work was a pair of
papers by Scherrer, Quashnock, Spergel and Press [12], and by Quashnock
and Spergel [13], which developed numerical and analytic techniques to
study the effects of gravitational back-reaction on the shape and
motion of the cosmic string loops.  This is the first work which
examines the way in which the shape of a string loop is changed as a
result of the emission of gravitational radiation.  (In our paper,
these effects are {\it not} taken into account - we assume periodic
motion of the loop).  In addition to verifying some of Burden's
results, they also obtained interesting results concerning the
distribution of $\gamma$ for typical families of non-self-intersecting
string loops.  Recent work by Allen and Shellard [14] used FFT (Fast
Fourier Transform) methods to determine values of $\gamma$ for the
loops produced in their numerical simulation of cosmic string networks
in an expanding universe.

These investigations are important for the reason mentioned previously;
the cosmological consequences of cosmic strings are largely visible via
the direct and indirect effects of the gravitational waves produced by
the string loops.  Thus, ``typical" or expected values of $\gamma$
appear in expressions for observable quantities such as the present-day
energy-density expected in gravitational waves.  Much of the research
work on gravitational radiation by cosmic string loops has been
motivated by a desire to determine the ``typical" or ``expected" values
of $\gamma$.  Thus, Scherrer, Quashnock, Spergel and Press [12] give a
histogram of the expected values of $\gamma$; the mean is $\gamma
=61.7$ and the median is 55.4.

In much of the literature on this topic, the method used to determine
$\gamma$ is numerical.  A loop of cosmic string radiates at discrete
frequencies corresponding to the different normal modes of motion, so
$\gamma=\sum_{n=1}^\infty \gamma_n$ is a sum of terms arising from each
of these normal modes, labeled by $n=1,2,3, \cdots$.   The value of
each $\gamma_n$ is given by an integral over the two-sphere of a
particular function.  This function, in turn, is a product of integral
transforms over the world-sheet of the loop.  Except in certain
highly-symmetric cases, numerical methods must be used to determine the
required four-dimensional integrals.  Because it is only practical to
determine $\gamma_n$ up to $n$ of a few hundred or thousand, one must
extrapolate the dependence on $n$ in order to estimate the ``tail"
terms arising in the infinite sum over $n$.  This process is
error-prone because the sum over $n$ may converge very slowly (if at
all - with back-reaction neglected, $\gamma$ may be infinite!).  Also,
if the integration over the two-sphere is not done accurately enough,
the $\gamma_n$ will be inaccurate for large $n$.  This will cause the
sum over $n$ to converge at the wrong rate.  Indeed, we have found that
many of the previously-determined values of $\gamma$ given in the
literature are incorrect (typically by a factor of order two)  because
this tail has either not been included, or has been incorrectly
estimated.

In this paper, we develop a new method for determining $\gamma$.   Our
method yields an {\it exact analytic formula for $\gamma$}, valid for
any piecewise linear cosmic string loop with piecewise linear
velocity.  (Equivalently, both the left- and right-moving trajectories
are piecewise linear).  This piecewise linear requirement is really not
very restrictive, since in practice any cosmic string loop can be {\it
arbitrarily closely approximated} by a piecewise linear cosmic string.
Thus {\it  one can use this formula to determine $\gamma$ to arbitrary
precision for any cosmic string loop.} Remarkably, our formula involves
nothing more complicated than log and arctangent functions.  However it
is the sum of order $N^{4}$ terms, where $N$ is the number of piecewise
linear segments, and thus in practice is extremely cumbersome to
evaluate without the assistance of a computer or symbolic manipulator.
We stress that although our formula will probably never be evaluated
without the use of a computer, it is {\it not a numerical method}, but
rather is an {\it exact formula}.  It is also fairly rapid - with $N$
of 32 a DEC ALPHA class workstation can evaluate $\gamma$ in less than
2.5 seconds.  We are making our computer code, which provides one
implementation of this algorithm, publicly available via anonymous FTP
from the directory pub/pcasper at alpha1.csd.uwm.edu.

In order to test our new formula, we constructed piecewise linear
approximations to the smooth cosmic string loops studied in earlier
published calculations of $\gamma$.   In a number of cases we obtained
very close agreement between the value of $\gamma$ given by our formula
and the published values.  However there were also a number of cases in
which the results did not agree.  Section 7 contains further details of
these cases.  In every case where we had found disagreement, we were
able to show that our formula in fact had given the correct result.
The disagreement in each case was due to numerical errors
in the original work.  Many of the published values of $\gamma$ are off
by about a factor of two.  For example, Vachaspati and Vilenkin give
the value $\gamma=54.0$ for the case $\alpha=0.5$ and $\phi=0.5\pi$ in
equation (2.24) of reference [9].  The correct value is $\gamma=97.2
\pm 2$.  Note that the value of $\gamma=97.2$ is exact (to three
significant figures) for the piecewise loop which we used to
approximate the smooth Vachaspati and Vilenkin loop.  The error bar of
$\pm 2$ in $\gamma$ arises because our piecewise approximation had only
$N=64$ segments.

We intend to use this exact formula in future work, for example to
identify the shape of a cosmic string loop with the smallest value
of $\gamma$, and to repeat some of the work of Scherrer, Quashnock,
Spergel and Press [12] concerning the distribution of $\gamma$
values of non-self-intersecting loops.

The remainder of the paper is organized as follows.  Section 2
describes the periodic motion of a cosmic string loop oscillating in
flat space-time.  It establishes notational conventions and a number
of basic results.  Note that in our approximation, the back-reaction of
gravity on the string loop is neglected, so that space-time remains
flat.  In this context a given string loop oscillates periodically and
radiates forever.  Section 3 starts with a standard result [15] for the
energy radiated by gravitational waves
emitted from a periodic source, and obtains an integral representation
for $\gamma$ in terms of the gravitational interaction of the cosmic
string world-sheet with itself.  This was motivated by (and is almost
identical to) a calculation given in appendix B of [13].  In Section
4 we restrict our attention to the special case
of piecewise linear loops, and establish notational conventions for
such loops.  The ``corners" of the piecewise linear loop trajectory may
be discretely labeled; their positions/velocities contain all
information about the loop.  The integral representation for $\gamma$
is then expressed in terms of these discrete quantities.  In Section 5,
the formula for $\gamma$ is simplified and expressed as a sum of
elementary integrals.  These integrals are three-dimensional volume
integrals; the integrand is a Dirac delta function of a quadratic form
in $x,y$ and $z$.  These integrals are
evaluated in closed form in Section 6.  This section contains the main
result of the paper, which is an exact closed-form expression for
$\gamma$ in the piecewise linear case.   Section 7 contains the results
of our investigation of the existing literature, reporting both on
those cases where we obtained agreement with published work, and those
cases where we found the published work to be incorrect.  In the latter
cases, we have isolated the error(s) in the published work and report
on how we corrected those errors.  Section 8 contains a short
``catalog" of values of $\gamma$ for some elementary loop
trajectories.  This is followed by a short conclusion.

Note: throughout this paper we use the metric
signature $(-,+,+,+)$, and denote Newton's constant by $G$.  From here
on we use units with the speed of light $c=1$.
\vskip0.3in
\centerline{Section 2: COSMIC STRING MOTION IN FLAT SPACE}
\vskip0.05in
The trajectory of a cosmic string describes a two-dimensional
world-sheet in space-time.
Points on the world-sheet have space-time coordinates $x^\mu$ given by
$$x^\mu=x^\mu(\xi^0,\xi^1),
\eqno(2.1)$$
where $\xi^0$ is a timelike and $\xi^1$ is a spacelike coordinate on the
world-sheet.  The string is described
by the Nambu action, which is proportional to the area of the world-sheet:
$$S=-\mu\int [-g^{(2)}]^{1/2}d^2\xi .
\eqno(2.2)$$
Here $\mu$ is the mass-per-unit length of the string, $g^{(2)}$ is the
determinant of the 2-dimensional metric on the world-sheet induced from
the Minkowski metric, and the integration is
over the entire world-sheet of the string.  If we define $x^{\mu},{}_a
={\partial x^{\mu}\over\partial\xi^a}$, where $a=0,1$, then the induced
2-dimensional metric is given by
$$g^{(2)}_{ab}=g_{\mu\nu}x^\mu,{}_ax^\nu,{}_b.
\eqno(2.3)$$
If we denote time and space derivatives on the world-sheet by
$\dot{}={\partial\over\partial\xi^0}$ and $'={\partial\over
\partial \xi^1}$, then the determinant $g^{(2)}$ is
$$g^{(2)}=\dot x^\mu\dot x_\mu x'{}^\nu x'_\nu-x'{}^\mu \dot
x_\mu\dot x^\nu x'{}_\nu.
\eqno(2.4)$$
Note that $x'{}^\mu$ is spacelike and $\dot x^\mu$ is timelike.

The Lagrangian equations of motion for the string are rather cumbersome [3].
However, the action (2.2) is invariant under the
reparameterization (gauge transformation) $\xi^a\rightarrow\tilde\xi^a(\xi)$,
 so the equations can be simplified by a judicious choice of the parameters
$\xi^a$.  One may choose the
parameters so that $x^\mu$ satisfies the gauge conditions
$$\dot x^\mu {x'}_\mu=0 \quad {\rm and} \quad\dot x^\mu\dot
x_\mu+{x'}^\mu {x'}_\mu=0.
\eqno(2.5)$$
With this choice of gauge, the equation of motion is the two-dimensional
wave equation
$$\ddot x^\mu-x''{}^\mu=0.
\eqno(2.6)$$
The gauge conditions (2.5) still allow a further reparameterization where
$\dot{\tilde \xi}{}^1=\tilde\xi'{}^0$ and
$\tilde \xi'{}^1=\dot{\tilde\xi}{}^0$.
Together these imply that $\ddot{\tilde\xi}{}^a=\tilde\xi''{}^a$.  This allows
us to set $\xi^0=t$.  If we rename $\xi^1=\sigma$, then the coordinates of
the string world-sheet (2.1) become
$$x^\mu=x^\mu(t,\sigma).
\eqno(2.7)$$
With this choice of parameters, the gauge conditions (2.5) become
$$\dot x^ix'_i=0, \quad { \rm and} \quad \dot x^i\dot x_i+x'{}^ix'_i=1,
\eqno(2.8)$$
where the index $i=1,2,3$ is a spatial index.  The equation of motion
becomes the two-dimensional wave equation
$$\ddot x^i-x''{}^i=0.
\eqno(2.9)$$
The time part of the equation of motion (2.6) is satisfied automatically.

The general solution to the equation of motion (2.9) is
$$\vec x(\sigma ,t)={1\over 2}[\vec a(t+\sigma)+\vec b(t-\sigma)].
\eqno(2.10)$$
Here, the function $\vec a$ defines the left-moving
and $\vec b$ the right-moving component of the
string.  The first gauge condition applied to $\vec x$ implies that
$\vec a'{}^2=\vec b'{}^2$, where here the prime means differentiation with
respect to the function's argument.  The second gauge condition implies that
$\vec a'{}^2+\vec b'{}^2=2$.  Together the gauge conditions force
the functions $\vec a$ and $\vec b$ to satisfy
$$\vec a'{}^2=\vec b'{}^2=1.
\eqno(2.11)$$
Up to this point, our treatment of cosmic strings includes both the
case of infinite strings and the case of closed string loops.  From
here on, to study gravitational radiation, we consider only the case of
closed loops.  In this case, the world-sheet of the (assumed
non-self-intersecting) string has the topology of a cylinder
$I\!\!R\times S^1$, and will be referred to as a ``world-tube".
Because the string forms a closed loop, one finds an additional
constraint on the otherwise arbitrary functions $\vec a$ and $\vec b$.

If the cosmic string has the form of a closed loop, it follows that
$$\vec x(t,\sigma+L)=\vec x(t,\sigma)\quad \forall\> \sigma,t,
\eqno(2.12)$$
where the constant $L$ is the length of the loop.  This implies that
$$\vec a(t+\sigma)+\vec b(t-\sigma)=\vec a(t+\sigma+L)+
\vec b(t-\sigma-L)\quad\forall\> \sigma, t.
\eqno(2.13)$$
If we define the null coordinates $u$ and $v$ by
$$u=t+\sigma,\quad v=t-\sigma,
\eqno(2.14)$$
then (2.13) becomes
$$\vec a(u+L)-\vec a(u)=\vec b(v)-\vec b(v-L)\quad\forall\> u,v.
\eqno(2.15)$$
However, because $u$ and $v$ can be varied independently, it must be
the case that
$$\vec a(u+L)-\vec a(u)=\vec b(v)-\vec b(v-L)=\vec c,
\eqno(2.16)$$
where $\vec c$ is a constant vector.  If we choose to work in the
center-of-mass frame of the loop, then $\vec c=0$.  This follows
since in the center-of-mass frame we have
$$\eqalign{0=\int\limits_0^L\dot {\vec x}d\sigma&=\int\limits_0^L{1\over
2}[\vec a'(t+\sigma)+\vec b'(t-\sigma)]d\sigma\cr&={1\over 2}[\vec a(t+L)
-\vec b(t-L)-\vec a(t)+\vec b(t)]=\vec c.\cr}
\eqno(2.17)$$
Thus, in the center-of-mass frame, the functions $\vec a$ and $\vec b$
are periodic with period $L$,
$$\vec a(t+\sigma+L)=\vec a(t+\sigma),\quad\vec b(t-\sigma-L)=
\vec b(t-\sigma).
\eqno(2.18)$$
Because the functions $\vec a$ and $\vec b$ are periodic, each can be
described by a closed loop.  These loops will be referred to
respectively as the a-loop and the b-loop.  Together, the a- and
b-loops define the trajectory of the string loop.

Because the functions $\vec a$ and $\vec b$ are periodic in their
arguments, the string loop is periodic in time.  The period of the
loop is $L/2$ since
$$\vec x(t+{L\over 2},\sigma+{L\over 2})={1\over 2}[\vec
a(t+\sigma+L)+\vec b(t-\sigma)]={1\over 2}[\vec a(t+\sigma)+
\vec b(t-\sigma)]=\vec x(t,\sigma).
\eqno(2.19)$$
For the remainder of this paper, we will set the loop length $L=1$.
The period of the loop is then $\Delta t=L/2=1/2$, and the section of
the world- tube swept out by the loop in a single oscillation is
covered by the coordinates $\sigma\in [0,1]$ and $t\in [0,1/2]$.  The
entire world- tube is covered by $\sigma\in [0,1]$ and $t\in
(-\infty,\infty)$.

The reason that one may set $L=1$ is remarkable: the power radiated in
gravitational radiation from a loop of a given shape is invariant under
a rescaling (magnification or shrinking) of the loop, provided that the
velocity at each point on the rescaled loop is unchanged [3].  A formal
proof of this is given in [14].  Thus, to calculate the radiated power
it is sufficient to consider only those loops with total length $L=1$.

The null coordinates $u$ and $v$ defined in (2.14) are more convenient
than the coordinates $t$ and $\sigma$.  The $u,v$ coordinates are called
null because the tangent four-vectors $\partial_u x^\alpha$ and
$\partial_v x^\alpha$ associated with them are null.  This follows because
$$\eqalign{{\partial\over\partial u}&={\partial\sigma\over\partial u}
{\partial\over\partial\sigma}+{\partial t\over\partial u}
{\partial\over\partial t}={\partial\over\partial\sigma}+
{\partial\over\partial t},\quad{\rm and}\cr\cr {\partial\over\partial v}
&={\partial\sigma\over\partial v} {\partial\over\partial\sigma}+
{\partial t\over\partial v} {\partial\over\partial t}=
-{\partial\over\partial\sigma}+{\partial\over\partial t},\cr}
\eqno(2.20)$$
are null vectors in Minkowski space.  To be more explicit,
since points on the world-tube have space-time coordinates
$[t,\vec x]$ given in terms of $u$ and $ v$ by
$$x^\a(u,v)={1\over 2}[u+v,\>\vec a(u)+\vec b(v)],
\eqno(2.21)$$
the gauge conditions now imply that
$$\eqalign{\partial_ux^\a\partial_ux_\a&=-\bigg({1\over 2}\bigg)^2+
\bigg({1\over 2}\bigg)^2 \>\vec a'{}^2=0,\quad\rm
and\cr\partial_vx^\a\partial_vx_\a&=-\bigg({1\over
2}\bigg)^2+\bigg({1\over2}\bigg)^2\>\vec b'{}^2=0.\cr}
\eqno(2.22)$$
Because of the periodicity of the loops, the world-tube may be covered
by the coordinates $u$ and $v$ in many equivalent ways.  One convenient
covering is to take $u\in[0,1]$ and $v\in (-\infty,\infty)$.  The
region of the world-tube swept out in a single oscillation of the loop
is then covered by $u\in[0,1]$ and $v\in[0,1]$.  Note however that this
is {\it not} the same region of the world-tube as $t\in[0,1/2]$ and
$\sigma\in[0,1]$.  This is shown in Figure 1 of reference [14].

The energy-momentum tensor $T^{\mu\nu}$ for the string loop may be
found by varying the action (2.2) with respect to the metric.  In flat
space, with our choice of coordinates and gauge, it is
$$T^{\mu\nu}(y^\a)=\mu\int\limits_0^1du\int\limits_{-\infty}^\infty dv\>
G^{\mu\nu}(u,v)\delta^4(y^\a-x^\a(u,v)),\eqno(2.23)$$
where $G^{\mu\nu}$ is defined by
$$G^{\mu\nu}(u,v)=\partial_ux^\mu\partial_vx^\nu+
\partial_vx^\mu\partial_ux^\nu.
\eqno(2.24)$$
Note that the volume element for the $(u,v)$ coordinates is related to
that of the coordinates $(t,\sigma)$ by the Jacobian of the coordinate
transformation.  Thus,
$$dudv=2d\sigma dt.
\eqno(2.25)$$
Because of the delta function which appears in (2.23),
the stress-tensor $T^{\mu\nu}$ vanishes everywhere except
on the world-sheet of the string loop.

\vfil
\eject
\vskip 0.3in \noindent
\centerline{Section 3:  POWER RADIATED IN GRAVITATIONAL RADIATION}
\vskip 0.05in
The power emitted by an oscillating loop in the form of gravitational
radiation may be determined in the weak-field limit.  This is an
excellent approximation for cosmologically interesting cosmic strings
because the amplitude of the metric perturbation $h_{\mu\nu}$ is of
order $G \mu/c^2 \approx 10^{-6}$.  Because the gravitational radiation
is weak, its back-reaction on the loops does not modify a loop's motion
significantly in a single oscillation.  Hence we calculate the rate of
gravitational radiation in the approximation that the back-reaction can
be neglected, so that a loop oscillates periodically in time.

The standard formulae used to calculate the power lost to gravitational
radiation typically assume that the energy of the source is gradually
dissipated into radiation, and that the stress-energy tensor of the
source vanishes with time.  In our case, the source is a periodically
oscillating loop whose stress-energy tensor does not vanish with time,
and therefore the standard formulae require minor modifications.
Since the loops that we study in this paper have period $1/2$, they
radiate only at discrete angular frequencies given by:
$$\omega_n = 4 \pi n \qquad {\rm for} \qquad n=1,2,3,\cdots .
\eqno(3.1)$$
The power radiated per unit solid-angle into the $nth$ mode is given
by the standard formula (equation 10.4.13 of [15])
$$ {dP_n \over d\Omega} = {G \over \pi} \omega_n^2 \bigl[
\tau^*_{\mu\nu}(\omega_n \Omega)
\tau^{\mu\nu}(\omega_n \Omega) - {1 \over 2}
|\tau^{\lambda}{}_{\lambda}(\omega_n \Omega) |^2 \bigr].
\eqno(3.2)$$
In this equation, the Fourier transform of the stress-energy tensor
is defined by
$$
\tau_{\mu\nu}(\omega \Omega) \equiv \lim_{T \to \infty}
{1 \over 2 T} \int_{-T}^{T} dt \int d^3x \>
e^{i \omega (t-  \Omega \cdot \vec x)} T_{\mu\nu}(t,\vec x),
\eqno(3.3)$$
and $*$ denotes complex conjugation.  The vector symbol over $\vec x$
denotes an ordinary flat-space three-vector, and $\Omega$ is a
unit-length three-vector with spatial Cartesian components given by
$$\Omega=(\cos\phi\sin\theta,\sin\phi\sin\theta,\cos\theta).
\eqno(3.4)$$
To calculate the total radiated power, one must integrate over
all directions on the unit sphere.  This introduces integrals of
the form
$$\int d\Omega f(\Omega)\equiv \int_0^\pi \sin \theta
d\theta \int_0^{2\pi} d\phi f(\theta,\phi)
\eqno(3.5)$$
into the equations that follow.

The Fourier transform of the stress-energy tensor (3.3) is defined as
the limit of a infinite-time integral.  This differs slightly from the
case of non-periodic sources.   It is easy to see that
$\tau_{\mu\nu}(\omega \Omega)$ vanishes unless $\omega$ takes on one of
the discrete values $\omega_n$.  (This is shown in reference [14]
following equation (3.11)).  For these values of $\omega_n$, the $T \to
\infty$ limit of the integral is not hard to calculate: since the
source is periodic, the limit appearing in (3.3) is equal to the
integral over a single period, i.e., the same integral with $2T$ set
equal to $1/2$.  Substituting expression (2.23) for the stress-energy
tensor into the formula for $\tau_{\mu\nu}$ one finds that
$$\tau_{\mu\nu}(\omega_n \Omega) =2\mu \idu \idv G_{\mu\nu}(u,v)
e^{i \omega_n (u+v - \Omega \cdot (\vec a(u) + \vec b(v)))/2}.
\eqno(3.6)$$
Note that the limits of integration in $(u,v)$ space cover one complete
oscillation period of the world-sheet of the string loop.

The total radiated power is obtained by summing over all the modes:
$$P =  \sum_{n=0}^\infty  \int d\Omega {dP_n \over d\Omega}.
\eqno(3.7)$$
The $n=0$ mode has been included in the sum for later convenience; it
makes no contribution because of the factor of $\omega_n^2$ appearing
in (3.2).  This expression for the total power can be put into a more
useful form by using the explicit formula (3.6) for the Fourier
transform of the stress-tensor. This gives
$$P= {2 G \mu^2 \over \pi} \sum_{n=-\infty}^{\infty}
 \omega_n^2 \int d\Omega
\idu \idv \idut \idvt
\psi(u,v,\ut,\vt)e^{i \omega_n (\Delta t(u,v,\ut,\vt) - \Omega \cdot \Delta
\vec x(u,v,\ut,\vt)),}
\eqno(3.8)$$
where we have defined
$$\psi(u,v,\ut,\vt)= G_{\mu\nu}(u,v) G^{\mu\nu}( \ut, \vt) -
{1 \over 2} G^\lambda{}_\lambda(u,v) G^\gamma{}_\gamma ( \ut ,\vt).
\eqno(3.9)$$
The functions $\Delta t = (u+v- \ut-\vt)/2$ and
$\Delta \vec x = (\vec a(u) + \vec b(v)-\vec a(\ut) - \vec b(\vt))/2$
in (3.8) describe the temporal and spatial separation of the two points on
the string world-sheet with coordinates $(u,v)$ and $(\ut,\vt)$ respectively.
To save space, in some of the formulae that follow, the arguments of
$\Delta t$ and $\Delta \vec x$ are not shown; they should be implicitly
understood.  Since each term in the sum over $n$ equals its complex-conjugate,
$P$ is explicitly real.  For this reason, since the $n=0$ term does not
contribute, we have changed the sum over $n$ to a sum from $-\infty$ to
$\infty$ at the expense of introducing an overall factor of $1/2$ into the
formula.  From here on, this sum will simply be denoted by $\sum_n$.

It is possible to carry out both the sum over $n$ and the
integral over solid angle in closed form.
To see this, consider the integral
$$I(t,\vec r) \equiv \sum_n \omega_n \int d\Omega
e^{i \omega_n (t - \Omega \cdot \vec r)}.\eqno(3.10)$$
The integral over solid angle $\int d\Omega
e^{i \Omega \cdot \vec z}$ is easily evaluated,
and equals $4 \pi \sin(|\vec z|)/|\vec z|$.
Hence one has
$$I (t,\vec r)={4 \pi\over |\vec r|}
 \sum_n  e^{i \omega_n t} \sin \omega_n |\vec r|.\eqno(3.11)$$
Because of the absolute value signs that appear, some care is
required to obtain this last result - one must separately consider
both possible signs of $\omega_n$.  The sum over $n$ may now be explicitly
carried out.  Using the standard formula
$\sum_n e^{i n \theta} = 2 \pi \delta_p(\theta)$ for the periodic delta
function on the interval $[- \pi,\pi)$, one obtains
$$I  (t,\vec r)= {4 \pi^2 \over i |\vec r|}
\bigl( \delta_p(4\pi(t + |\vec r|))
- \delta_p(4 \pi(t-|\vec r|)) \bigr).\eqno(3.12)$$
Noting further that the periodic delta function may be expressed in terms
of the ordinary Dirac delta function as
$$ \delta_p(x) = \sum_{k=-\infty}^{\infty} \delta(x+2 \pi k),\eqno(3.13)$$
and noting that $|\vec r|$ is always positive, one may
combine the two delta functions in (3.12) to give
$$\eqalign{I  (t,\vec r) &=
{\pi\over i|\vec r|} \sum_{k=-\infty}^{\infty}
\bigl[ \delta(t+k/2+|\vec r|) - \delta(t+k/2-|\vec r|) \bigr] \cr
&= 2 \pi i \sum_{k=-\infty}^{\infty}
\epsilon(t+k/2) \delta((t+k/2)^2 - |\vec r|^2),\cr}\eqno(3.14)$$
where $\epsilon (x)=2\theta (x)-1$ is $+1$ for $x>0$ and $-1$
for $x<0$.  From the definition (3.10) of $I(t,\vec r)$ it is clear that
applying the derivative operator $-i \partial/\partial t$
brings down an additional factor of $\omega_n$.
Inserting the time-derivative of (3.14) into (3.8) leads to
$$P= 4 G \mu^2 \sum_{k=-\infty}^{\infty}
\idu \idv \idut \idvt
\psi(u,v,\ut,\vt)
{\partial \over \partial \Delta t} \epsilon( \Delta t + k/2)
\delta( (\Delta t + k/2)^2-|\Delta \vec x|^2).\eqno(3.15)
$$
In this expression, the dependence of $\Delta t$ and $\Delta \vec x$ on
the four variables $u,v,\ut$ and $\vt$ has not been explicitly
shown. The derivative operator
$\partial \over \partial \Delta t$ means:
{\it first} take the derivative of $I(t,\vec r)$ with respect to
the first argument, {\it then} substitute in the functions
$\Delta t$ and $\Delta \vec x$.  Alternatively, it refers to any
combination of derivative operators in (3.8) which will bring down
a factor of $i \omega_n$.

It is possible to re-express the sum over $k$ of these four integrals as
a single integral, simply by shifting one of the  integration variables
to the range $-\infty$ to $\infty$.
For example, if we choose to shift $\vt$, then because the functions
$\vec a$ and $\vec b$ are periodic, it is easy to see that
$\Delta \vec x(u,v,\ut,\vt-k)=\Delta \vec x(u,v,\ut,\vt)$
and that $\psi(u,v,\ut,\vt-k) = \psi(u,v,\ut,\vt)$.
However, the time function is {\it not} periodic; one has
$\Delta t(u,v,\ut,\vt-k)=\Delta t(u,v,\ut,\vt)+k/2$.
Since the period of the loop is $1/2$, the energy radiated in a single
oscillation of the loop is thus given by $E=P/2$:
$$E= 2 G \mu^2
\idu \idv \idut \int_{-\infty}^{\infty} d\vt\>
\psi(u,v,\ut,\vt)
{\partial \over \partial \Delta t} \epsilon( \Delta t)
\delta( (\Delta t)^2-|\Delta \vec x|^2).\eqno(3.16)$$
Note that the choice to shift $\vt$ was arbitrary; we could
have chosen to shift any one of the four integration variables.
If we had chosen to shift some other variable, the only changes to
(3.16) would be that the integration range for $\vt$ would be from
0 to 1, and the integration range of the new shifted variable would
be from $-\infty$ to $\infty$.
This expression for the energy radiated into gravitational radiation
during one oscillation of the cosmic-string loop can be evaluated
exactly in the case of piecewise linear string loops.

To make this formula directly useful, one must replace the
operation $\partial/\partial \Delta t$ by an explicit operation
in terms of derivatives w.r.t. the
variables $u,v,\ut$ and $\vt$.  The
desired effect of this operation is to bring down a factor of
$i \omega_n$ when applied to
$ \exp ({i \omega_n [\Delta t(u,v,\ut,\vt) - \Omega \cdot \Delta
x(u,v,\ut,\vt)]})$.
Let us denote this operation by
\def\Ut{{\tilde U}}
\def\Vt{{\tilde V}}
$$ D(u,v,\ut,\vt) \equiv
U(u,v,\ut,\vt) \partial_u+V(u,v,\ut,\vt) \partial_v -
\Ut (u,v,\ut,\vt) \partial_{\ut} -
\Vt (u,v,\ut,\vt) \partial_{\vt},\eqno(3.17)$$
where the functions $U,V,\Ut$, and $\Vt$ are determined by
the desired effect of $D$ on the exponential:
$$ D \exp ({i \omega_n [\Delta t - \Omega \cdot \Delta \vec
x]}) = i \omega_n \exp ({i \omega_n [\Delta t - \Omega \cdot \Delta \vec
x]}) .\eqno(3.18)$$
Because $D$ is chosen to be a linear differential operator,
(3.18) is equivalent to the four equations
$$ D \Delta t(u,v,\ut,\vt) = 1 \qquad {\rm and}
\qquad D \Delta \vec x (u,v,\ut,\vt) = \vec 0.\eqno(3.19)$$
Substituting in the definitions of $\Delta t$ and
$\Delta \vec x$, this may be written as the $4 \times 4$ matrix equation
$$
\pmatrix{
1 & 1 & 1 & 1 \cr
\vec a'(u) & \vec b'(v) & \vec a'(\ut) & \vec b'(\vt) \cr}
\pmatrix{
U \cr V \cr \Ut \cr \Vt \cr}
=
\pmatrix{
2 \cr
\vec 0 \cr}.\eqno(3.20)$$
The solution to these linear equations yields the
following expression for the differential operator D:
\def\nnx{{\vec a'(u)}}
\def\nny{{\vec b'(v)}}
\def\nnz{{\vec a'(\ut)}}
\def\nnt{{\vec b'(\vt)}}
\def\ccx{{\nny \cdot [\nnz \times \nnt]}}
\def\ccy{{\nnx \cdot [\nnz \times \nnt]}}
\def\ccz{{\nnx \cdot [\nny \times \nnt]}}
\def\cct{{\nnx \cdot [\nny \times \nnz]}}
$$\eqalign{D&(u,v,\ut,\vt) = \cr
&2 { \ccx \partial_u - \ccy \partial_v -
\ccz \partial_\ut + \cct \partial_\vt
\over
\ccx-\ccy+\ccz-\cct}.\cr}\eqno(3.21)$$
In terms of the operator $D$, the energy radiated in gravitational waves
during one oscillation of the string loop is given by
$$E= 2 G \mu^2
\idu \idv \idut \int_{-\infty}^{\infty} d\vt\>
\psi(u,v,\ut,\vt)
D(u,v,\ut,\vt)\biggl( \epsilon( \Delta t)
\delta( (\Delta t)^2-|\Delta \vec x|^2)\biggr) .\eqno(3.22)$$
In this expression
the differential operator $D$ acts on all the quantities that stand
to its right.

Before continuing, we note that half the
terms may be easily eliminated from (3.22).  Under the operation
of interchanging the variables
$(u,v)$ with $(\ut,\vt)$, $U(\ut,\vt,u,v)=\Ut(u,v,\ut,\vt)$,
$V(\ut,\vt,u,v)=\Vt(u,v,\ut,\vt)$ and $\Delta t(\ut,\vt,u,v)=-\Delta
t(u,v,\ut,\vt)$, while $\psi(\ut,\vt,u,v)=\psi(u,v,\ut,\vt)$ and the
arguments of the $\delta$ function are invariant.  Using these results,
we interchange the variables $(u,v)$ with $(\ut,\vt)$ in the first two
terms of $E$ in (3.22).  Recalling that we could have shifted any one
of the integration variables in (3.15), as explained following (3.16),
one finds that
$$\eqalign{&\idu \idv \idut \int_{-\infty}^{\infty} d\vt\>
\psi(u,v,\ut,\vt)\big(U\partial_u+V\partial_v\big)\biggl(
\epsilon( \Delta t)\delta( (\Delta t)^2-|\Delta \vec x|^2)\biggr)\cr
\noalign{\vskip2pt}
&=\idu \idv \idut \int_{-\infty}^{\infty} d\vt\>
\psi(u,v,\ut,\vt)\big(\Ut\partial_\ut+\Vt\partial_\vt\big)\biggl(
\epsilon(-\Delta t)\delta( (\Delta t)^2-|\Delta \vec x|^2)\biggr).\cr}
\eqno(3.23)$$
This equation, along with $\epsilon (-x)=-\epsilon (x)$, may now
be used in (3.22) to eliminate two of the
four terms in the operator $D$, yielding
$$E= 4 G \mu^2 \idu \idv \idut \int_{-\infty}^{\infty} d\vt\>
\psi(u,v,\ut,\vt)\big(U\partial_u+V\partial_v\big)\biggl(
\epsilon( \Delta t)\delta( (\Delta t)^2-|\Delta \vec x|^2)\biggr) .
\eqno(3.24)$$
Although (3.24) has only half as many terms as (3.22), it is not the
most useful form for our purposes.

The most convenient expression for $E$ is obtained by replacing
$\epsilon (\Delta t)$ in (3.22) by $2\theta(\Delta t)$.  The
$\epsilon(\Delta t)=2 \theta(\Delta t) - 1$ term can be replaced by $2
\theta(\Delta t)$ because the $-1$ term makes no contribution to the
integral.  To see this, consider the effect of replacing
$\epsilon(\Delta t)$ by $-1$ in (3.22).  Denoting this by $E_{(-1)}$
and again using the transformation properties of $U(u,v,\ut,\vt)$,
$V(u,v,\ut,\vt)$, $\psi(u,v,\ut,\vt)$ and the arguments of the $\delta$
function under the operation of interchanging the variables $(u,v)$
with $(\ut,\vt)$, one finds that
$$\eqalign{E_{(-1)}\equiv & -2 G \mu^2
\idu \idv \idut \int_{-\infty}^{\infty} d\vt\> \psi(u,v,\ut,\vt)
\biggl(U\partial_u+V\partial_v -
\Ut\partial_{\ut} -\Vt \partial_{\vt}
\biggr)\delta( (\Delta t)^2-|\Delta \vec x|^2)\cr
\noalign{\vskip2pt}
=&-2 G \mu^2
\idu \idv \idut \int_{-\infty}^{\infty} d\vt\> \psi(u,v,\ut,\vt)
\biggl(\Ut\partial_{\ut}+\Vt\partial_{\vt} -
\Ut\partial_{\ut} -\Vt \partial_{\vt}
\biggr)\delta( (\Delta t)^2-|\Delta \vec x|^2)=0.\cr}\eqno(3.25)$$
This proves that $E_{(-1)}=0$ and thus that $\epsilon(\Delta t)$ in
(3.22) can be replaced by $2 \theta(\Delta t)$.

The integration range $\vt\in (-\infty,\infty)$ in the expression for
$E$ (3.22) may be replaced by the finite range $\vt \in [-2,2]$.  This
is because the integrand of (3.22) vanishes unless $\vt \in [-2,2]$.
Physically this is because in the center-of-mass frame, the string loop
remains centered about a fixed coordinate location.   Since the
$\delta$ function has its support only on the light-cone, it is
impossible for regions of the string located far in the past or future
to interact.  To see this result mathematically, first consider the
argument of the $\theta$ function.  The only contributions to $E$ arise
when this argument is positive, i.e., when
$$ u+v - \ut - \vt \ge 0. \eqno(3.26)$$
Since all three integration variables $u,v,\ut$ lie
in the range $[0,1]$, the $\theta$-function vanishes unless the
variable $\vt$ lies in the range $\vt \in (-\infty,2]$.  Further
restrictions arise from considering the argument of
the $\delta$ function.   The only contributions to
$E$ arise when this argument vanishes, i.e., when
$$(u+v-\ut - \vt)^2 = [\vec a(u) - \vec a (\ut) +
\vec b(v) - \vec b(\vt)]^2.
\eqno(3.27)$$
However since the total length of the $a$-loop is
$1$, the maximum length of the vector $\vec a(u) - \vec
a(\ut)$ is $1/2$.  Similarly, the maximum length of
the vector $\vec b(v) - \vec b(\vt)$ is $1/2$.  Hence
the largest possible value of the r.h.s. of (3.27)
is $1$.   This shows that the integrand vanishes unless
the quantity
$$ u+v - \ut - \vt \in [-1,1].
\eqno(3.28)$$
Again making use of the possible ranges of $u$, $v$,
and $\ut$ this implies that the
$\delta$ function vanishes unless $\vt \in [-2,3]$.
Combined with the restrictions arising from the
$\theta$ function, this implies that the only
contributions to $E$ arise from the range $\vt \in [-
2,2]$.   Thus we obtain the final form of our result
$$E= 4 G \mu^2
\idu \idv \idut \int_{-2}^{2} d\vt\>
\psi(u,v,\ut,\vt)
D(u,v,\ut,\vt)\biggl( \theta( \Delta t)
\delta( (\Delta t)^2-|\Delta \vec x|^2)\biggr) .
\eqno(3.29)$$
The expression just obtained has an interesting physical
interpretation.

The string loses energy to gravitational waves precisely because of the
gravitational self-interaction of the string with itself.  From
this point of view, the integral over $(\ut,\vt)$ in (3.29)
is an integral over the sources (the stress-tensor of the string
worldsheet at $x^{\alpha}(\ut,\vt)$) that create a metric perturbation
at the space-time point $x^\alpha (u,v)$. The metric perturbation at
$x^\alpha (u,v)$ is obtained by multiplying the source times a retarded
propagator and integrating over the entire history of the source: the
part of the world-sheet that can contribute is covered by the
coordinates $\ut\in[0,1]$ and $\vt\in [-2,2]$.  The metric
perturbations from the loop at one space-time point propagate along the
light-cone from that point to interact with some other point on the
loop at some later time.  The product $\theta \delta/2 \pi$ that
appears in (3.29) is precisely the retarded propagator (equation 12.133
of [16]).  This creates the mechanism for energy loss:  the string must
do work against the tidal forces created by the metric perturbations
due to the string itself.  The energy lost in a single oscillation is
obtained by integrating this work over the region on the string's
world-tube covered by the coordinates $u\in[0,1]$ and $v\in[0,1]$.
Thus, the loss of energy due to gravitational radiation may be thought
of in terms of a loop which creates metric perturbations, interacting
with a loop whose motion does work against these perturbations.

We will see in the following sections that (3.29) can be evaluated {\it
exactly} in {\it closed analytic form} for any piecewise linear cosmic
string loop.  The value of $\gamma$ is then given immediately by
(1.1).

\def\api{{\vec a'_i}}

\def\bpj{{\vec b'_j}}

\def\apk{{\vec a'_k}}

\def\bpl{{\vec b'_l}}

\def\ai{{\vec a_i}}
\def\aii{{\vec a_{i+1}}}
\def\bj{{\vec b_j}}

\def\ak{{\vec a_k}}

\def\bl{{\vec b_l}}

\def\tpx{{\bpj \cdot [\apk \times \bpl]}}
\def\tpy{{\api \cdot [\apk \times \bpl]}}
\def\tpz{{\api \cdot [\bpj \times \bpl]}}
\def\tpt{{\api \cdot [\bpj \times \apk]}}

\vskip 0.3in \centerline{Section 4: PIECEWISE LINEAR LOOPS}
\vskip 0.05in
We now restrict our attention to piecewise linear loops.  These are
loops for which the functions $\vec a(u)$ and $\vec b(v)$, which
define the loop's trajectory, are piecewise linear functions.  The
functions $\vec a(u)$ and $\vec b(v)$ may be pictured as a pair
of closed loops, which consist of
joined straight {\it segments}.  The segments on the a- and b-loops
join together at {\it kinks} where $\vec a'(u)$ and $\vec b'(v)$ are
discontinuous.  The a-loop has $N_a$ linear segments, and the b-loop
has $N_b$ linear segments.  Part of a typical a-loop is shown in
Figure 1.

The following conventions, also shown in Figure 1, are used to describe
piecewise linear loops.  The coordinate $u$ on the a-loop is chosen
to take the value zero at one of the kinks, and increases along the
loop.  The {\it kinks} on the loop are labeled by the index $i$ where
$i=0,1,2,\dots,N_a-1$.  The value of $u$ at the $i$th kink is denoted
by $u_i$.  Without loss of generality we set $u_0=0$.  The {\it segments}
on the loop are also
labeled by the index $i$; the $i$th segment is the one lying between
the $i$th kink and the $(i+1)$th kink.  The kink at $u=u_{N_a}$ is
the same as the first kink at $u=u_0=0$.  Since $|\vec a'|^2=1$, the
a-loop has length 1.  Thus, even though $u_0$ and $u_{{N_a}}$ are at
the same position on the loop, $u_0=0$ while $u_{{N_a}}=1$.  The entire
range $u\in(-\infty, \infty)$ may be covered by allowing the coordinate
$u$ to continue around the a-loop in a periodic way.  This also extends
the range of the index to $i\in Z\!\!\!Z$.  Thus, for example,
$u_{-N_a}=-1, u_0=0, u_{N_a}=1, u_{2N_a}=2$ and so on are all located
at the same position on the a-loop.  Because $|\vec a'(u)|=1$, the length
of the linear segment between the kinks at $u_i$ and $u_{i+1}$ (the
$i$th segment) is
$$\Delta u_i=u_{i+1}-u_i.\eqno(4.1)$$
The loop's position $\vec a(u)$ at $u=u_i$ is denoted
$$\ai=\vec a (u_i).\eqno(4.2)$$
The constant unit vector tangent to the $i$th segment (pointing in the
direction of increasing $u$-parameter) is denoted
$$\api={\aii-\ai\over \Delta u_i}.\eqno(4.3)$$
With these definitions, the function $\vec a(u)$ for
$u\in[u_i,u_{i+1}]$ may be written
$$\vec a(u)=\ai+\api(u-u_i) \quad {\rm for} \> u\in[u_i,u_{i+1}].
\eqno(4.4)$$
Note that for consistency, putting $u=u_{i+1}$ in (4.4), one must
have
$$\aii=\ai+\api(\Delta u_i).\eqno(4.5)$$
Similar notation is used for the function $\vec b(v)$.  For
$v\in[v_j,v_{j+1}]$, the function $\vec b(v)$ may be written
$$\vec b(v)=\bj+\bpj(v-v_j) \quad {\rm for} \> v\in[v_j,v_{j+1}].
\eqno(4.6)$$
Thus the $a$-loop is entirely specified by the quantities
$\vec a_i$; from them one can obtain both
$\Delta u_i = |\aii - \ai|$ and $\api$ given by (4.3).
Identical notation is used for the $b$-loop.

The function $\psi(u,v,\ut,\vt)$ takes a special form in the case of
piecewise linear loops.  In general one has
$$\eqalign{\psi(u,v,\ut,\vt)&=
G^{\mu\nu}(u,v)G_{\mu\nu}(\ut,\vt)-{1\over 2} G^\lambda{}_\lambda(u,v)
G^\gamma{}_\gamma(\ut,\vt)\cr &=2[\partial_ux^\mu\partial_vx^\nu
\partial_\ut \xt_\mu\partial_\vt\xt_\nu+\partial_ux^\mu\partial_vx^\nu
\partial_\ut
\xt_\nu\partial_\vt\xt_\mu-\partial_ux^\lambda\partial_vx_\lambda
\partial_\ut \xt^\gamma\partial_\vt\xt_\gamma],\cr}\eqno(4.7)$$
where, again, $x^\a=x^\a(u,v)$ and $\xt^\a=x^\a(\ut,\vt)$.
For the purpose of evaluating integral (3.29) it is necessary to break
up the integrations over $(u,v,\ut,\vt)$ into rectangular four-cells.
Each four-cell is denoted by a set of indices $(i,j,k,l)$.  The
indices refer to segments on the a- and b-loop, each of which
defines a specific range for one of the coordinates $(u,v,\ut,\vt)$.
The index $i$ will always refer to the $u$ coordinate, $j$ to the $v$
coordinate, $k$ to the $\ut$ coordinate and $l$ to the $\vt$ coordinate.
Within each cell, we may write the four-vectors and tangent vectors
$$\matrix{x^\a(u,v)={1\over 2}[u+v,\ai+\api(u-u_i)+\bj+\bpj(v-v_j)]\cr
\noalign{\vskip4pt}
\partial_ux^\a={1\over 2}[1,\api],\quad \partial_vx^\a=
{1\over 2}[1,\bpj]}\Bigg\rbrace \quad {\rm for} \> u \in
[u_i,u_{i+1}] \> {\rm and} \> v \in [v_j,v_{j+1}]
\eqno(4.8)$$
and similarly
$$\matrix{x^\a(\ut,\vt)={1\over 2}[\ut+\vt,\ak+\apk(\ut-u_k)+
\bl+\bpl(\vt-v_l)]\cr
\noalign{\vskip4pt}
\partial_\ut\xt^\a={1\over 2}[1,\apk],\quad\partial_\vt\xt^\a=
{1\over 2}[1,\bpl]}\Bigg\rbrace \quad {\rm for}\> \ut \in [u_k,u_{k+1}]
\> {\rm and} \> \vt \in [v_l,v_{l+1}].
\eqno(4.9)$$
Using (4.8) and (4.9) in (4.7) we find that $\psi(u,v,\ut,\vt)$ is a constant,
$\psi_{ijkl}$, when $(u,v,\ut,\vt)$ are in the intervals defined by the
segments $(i,j,k,l)$. For any set of segments $(i,j,k,l)$, the constant
$\psi_{ijkl}$ is given by
$$\psi_{ijkl}={1\over 8}[(-1+\api\cdot\apk)(-1+\bpj\cdot\bpl)+
(-1+\api\cdot\bpl)(-1+\bpj\cdot\apk)-(-1+\api\cdot\bpj)(-1+\bpl\cdot\apk)].
\eqno(4.10)$$
Note that $\psi_{ijkl}\in[-1,5/4]$.  Also note that $\psi_{ijkl}$
vanishes when $(i-k)$ mod $N_a=0$ or $(j-l)$ mod $N_b=0$.

It is helpful to keep track of whether the indices in a given equation
refer to kinks or segments on the a- and b-loop.  For instance, $u_i$
refers to the value of the parameter $u$ at the $i$th kink on the
a-loop.  Similarly, $\vec a_i$ denotes a vector from the origin to the
kink at $u=u_i$ on the a-loop.   By contrast, $\vec a'_i$ is a unit
vector parallel to a specific segment on the a-loop.  The indices on
$\psi_{ijkl}$ also refer to specific segments on the a- and b-loops.
The segments define specific ranges for the coordinates
$(u,v,\ut,\vt)$.  Of course these ranges will change as the indices
take on different values.

The formula (3.29) for the energy radiated in gravitational waves
during one oscillation of the string loop may now be rewritten for the
case of a piecewise linear loop.  The integrals over $(u,v,\ut,\vt)$ in
(3.29) may be broken up into a sum of integrals over the individual
segments making up the a- and b-loops.  Because $\psi_{ijkl}$ is a
constant in each of these integrals, it may be pulled out of the
integration, giving
$$E=4G\mu^2\sum\limits_{i=0}^{{N_a}-1}\sum\limits_{j=0}^{{N_b}-1}
\sum\limits_{k=0}^{{N_a}-1}\sum\limits_{l=-2{N_b}}^{2{N_b}-1}\psi_{ijkl}
\int\limits_{u_i}^{u_{i+1}}du\int\limits_{v_j}^{v_{j+1}}dv
\int\limits_{u_k}^{u_{k+1}}d\ut\int\limits_{v_l}^{v_{l+1}}d\vt\>
D_{ijkl}(u,v,\ut,\vt)\big[\theta(\Delta t)\delta\big((\Delta t)^2-|\Delta \vec
x|^2\big)\big]. \eqno(4.11)$$
Note that the summation of $l$ is not from $-\infty$ to $\infty$ but
is only over the finite range corresponding to $\vt \in [-2,2]$ as shown
following equation (3.28).  Here
$$D_{ijkl}(u,v,\ut,\vt)=U_{ijkl}\partial_u+V_{ijkl}
\partial_v-\tilde U_{ijkl}\partial_{\ut}-\tilde
V_{ijkl}\partial_{\vt},\eqno(4.12)$$
where the coefficients of the derivative operators are constant on any
$(i,j,k,l)$ segment, and are given by
$$\eqalign{
U_{ijkl}= Q_{ijkl} \> \tpx \quad &\tilde U_{ijkl}=Q_{ijkl} \> \tpz \cr
V_{ijkl}= - Q_{ijkl} \> \tpy \quad &\tilde V_{ijkl}=-Q_{ijkl} \> \tpt ,\cr}
\eqno(4.13)$$
with (twice) the inverse determinant given by
$$Q_{ijkl} = 2 \biggl( \tpx - \tpy + \tpz - \tpt \biggr)^{-1}.\eqno(4.14)$$
Note that $U_{ijkl}$, $V_{ijkl}$,
$\tilde U_{ijkl}$ and $\tilde V_{ijkl}$ are all constant for a given set
$(i,j,k,l)$.  Again, note that the indices $(i,j,k,l)$ in
equations like (4.12) refer to specific straight
segments on the a- and b-loops.  They
do {\it not} refer to the components of some tensor.

\vskip 0.3in
\centerline{Section 5: EVALUATING THE INTEGRALS}
\vskip 0.05in
The four partial derivative operators in
$D_{ijkl}(u,v,\ut,\vt)$ in (4.11)
may be trivially integrated over
$u$, $v$, $\ut$ or $\vt$.  Carrying out
these integrations, $E$ takes the form
$$\eqalign{E=8G\mu^2 \sum\limits_{i=0}^{{N_a}-1}\sum\limits_{j=0}^{{N_b}-1}
\sum\limits_{k=0}^{{N_a}-1}\sum\limits_{l=-2{N_b}}^{2{N_b}-1}\psi_{ijkl}&
\big[U_{ijkl}\big(S^{(u)}_{i+1,j,k,l}-S^{(u)}_{i,j,k,l}\big)
+V_{ijkl}\big(S^{(v)}_{i,j+1,k,l}-S^{(v)}_{i,j,k,l}\big)\cr &-\tilde
U_{ijkl}\big(S^{(\ut)}_{i,j,k+1,l}-S^{(\ut)}_{i,j,k,l}\big)
-\tilde V_{ijkl}\big(S^{(\vt)}_{i,j,k,l+1}-S^{(\vt)}_{i,j,k,l}\big)
\big],\cr} \eqno(5.1)$$
where the superscript on each $S$ denotes the variable which has been
integrated out in that term.
The quantities $S^{(u)}_{ijkl}$, $S^{(v)}_{ijkl}$, $S^{(\ut)}_{ijkl}$ and
$S^{(\vt)}_{ijkl}$ appearing in (5.1) may all be
expressed in terms of a three-dimensional integral containing
a $\delta$ function:
$$\eqalign{S(\Delta x,\Delta y, \Delta z,& \tau, s, C(M,N,a,b,c,d),
\b_1, \b_2, \b_3, \b_4, \b_5, \b_6, \b_7)\cr
&=\int\limits_0^{\Delta x}dx \int\limits_0^{\Delta y}dy\int\limits_0^
{\Delta z}dz\>\theta(\tau+s(x+y-z))\delta
(\b_1xz+\b_2yz+\b_3xy+\b_4z+\b_5x+\b_6y+\b_7).\cr}\eqno(5.2)$$
Notice that the limits of integration
have been shifted so that the lower limit is always zero.
The function $C$ is defined by
$$\eqalign{
C( & M,N,a,b,c,d) =\cases{
1 & if  ${a-c-1\over M}={d-b\over N}=$\ integer \cr&\cr
2 & if  ${a-c-1\over M}={d-b-1\over N}=$\ integer \cr&\cr
3 & if  ${a-c+1\over M}={d-b-1\over N}=$\ integer \cr&\cr
4 & if  ${a-c+1\over M}={d-b\over N}=$\ integer \cr&\cr
0 &  otherwise. \cr}\cr}\eqno(5.3)$$
The function $S$ does not depend upon $C$, since $C$ does not appear
on the r.h.s. of (5.2).  However $C$ provides a convenient
means to later simplify certain special cases that arise.
The quantities $S^{(u)}_{ijkl}$, $S^{(v)}_{ijkl}$, $S^{(\ut)}_{ijkl}$ and
$S^{(\vt)}_{ijkl}$ are given in terms of $S$ by
$$\eqalign{S^{(u)}_{ijkl}=S(\Delta v_l,&\Delta u_k, \Delta v_j, M_{ijkl}, -1
, C(N_b,N_a,l,k,j,i), (1-\bpj\cdot\bpl), (1-\apk\cdot\bpj),
(\apk\cdot\bpl-1),\cr &
(\bpj\cdot\vec N_{ijkl}-M_{ijkl}), (M_{ijkl}-\bpl\cdot\vec N_{ijkl}),
 (M_{ijkl}-\apk\cdot\vec N_{ijkl}),
{1\over 2}(\vec N_{ijkl}^2-M_{ijkl}^2))\cr\cr
S^{(v)}_{ijkl}=S(\Delta u_k,&\Delta v_l, \Delta u_i, M_{ijkl}, -1,
C(N_a,N_b,k,l,i,j),
 (1-\api\cdot\apk), (1-\api\cdot\bpl), (\apk\cdot\bpl-1), \cr &
(\api\cdot\vec N_{ijkl}-M_{ijkl}), (M_{ijkl}-\apk\cdot\vec N_{ijkl}),
(M_{ijkl}-\bpl\cdot\vec N_{ijkl}),
{1\over 2}(\vec N_{ijkl}^2-M_{ijkl}^2))\cr\cr
S^{(\ut)}_{ijkl}=S(\Delta v_j,& \Delta u_i, \Delta v_l, M_{ijkl}, +1,
C(N_b,N_a,j,i,l,k)
, (1-\bpj\cdot\bpl), (1-\api\cdot\bpl), (\api\cdot\bpj-1), \cr &
(M_{ijkl}-\bpl\cdot\vec N_{ijkl}), (\bpj\cdot\vec N_{ijkl}-M_{ijkl}),
 (\api\cdot\vec N_{ijkl}-M_{ijkl}),
{1\over 2}(\vec N_{ijkl}^2-M_{ijkl}^2))\cr\cr
S^{(\vt)}_{ijkl}=S(\Delta u_i,& \Delta v_j, \Delta u_k, M_{ijkl}, +1,
C(N_a,N_b,i,j,k,l),
 (1-\api\cdot\apk), (1-\apk\cdot\bpj), (\api\cdot\bpj-1), \cr &
(M_{ijkl}-\apk\cdot\vec N_{ijkl}), (\api\cdot\vec N_{ijkl}-M_{ijkl}),
(\bpj\cdot\vec N_{ijkl}-M_{ijkl}),
{1\over 2}(\vec N_{ijkl}^2-M_{ijkl}^2)),\cr}
\eqno(5.4)$$
where we have defined
$$\eqalign{M_{ijkl}&=u_i+v_j-u_k-v_l,\quad {\rm and}\cr \vec N_{ijkl}
&=\ai+\bj-\ak-\bl.\cr}\eqno(5.5)$$
Note that the three remaining integrations in $S$ may be done in any order.
The relationships
$$\eqalign{&S(\Delta x,\Delta y,\Delta z,\tau,s,
\b_1,\b_2,\b_3,\b_4,\b_5,\b_6,\b_7)\cr
=&S(\Delta y,\Delta x,\Delta z,\tau,s,\b_2,\b_1,\b_3,\b_4,\b_6,\b_5,\b_7)\cr
=&S(\Delta x,-\Delta z,-\Delta y,\tau,s,-\b_3,\b_2,-\b_1,-\b_6,
\b_5,-\b_4,\b_7)\cr}
\eqno(5.6)$$
allow one to rewrite the integrations in any order.

It should be noted that in the definition (5.2) of $S$, there are no
terms in the argument of the $\delta$ function that are quadratic in
$x, y$ or $z$.  This is because the terms $u^2, v^2, \ut^2$ and $\vt^2$
in $\Delta t^2-|\Delta\vec x|^2$ appear  with respective coefficients
$$-{1\over 2}(1-\api\cdot\api),\quad -{1\over 2}(1-\bpj\cdot\bpj),\quad
-{1\over 2}(1-\apk\cdot\apk),\quad -{1\over 2}(1-\bpl\cdot\bpl).
\eqno(5.7)$$
It is easy to see that these coefficients all vanish since
$|\vec a'|^2=|\vec b'|^2=1$.  This is because, as described in section 2,
the coordinates $u$ and $v$ are null coordinates.
\vskip 0.3in
\centerline{Section 6: EVALUATION OF S}
\vskip 0.05in
The previous section reduced the problem of determining $\gamma$ in the
piecewise linear case to evaluating a set of integrals defined by the
function $S$ in equation (5.2).  In this section we carry out that
evaluation in closed form.  The $\delta$ function appearing in (5.2)
allows us to reduce the number of integrations in $S$ from three to
two.  This $\delta$ function may be written as $\delta(f(x,y,z))$,
where the argument of the $\delta$ function is
$$f(x,y,z)=\b_1xz+\b_2yz+\b_3xy+\b_4z+\b_5x+\b_6y+\b_7.\eqno(6.1)$$
The $\delta $ function will only have support when $f(x,y,z)=0$.
Solving $f(x,y,z)=0$ for $x(y,z)$, $y(x,z)$ and $z(x,y)$ respectively,
we find that
$$x(y,z)= - { \beta_2 y z + \beta_4 z + \beta_6 y + \beta_7 \over
              \beta_1 z + \beta_3 y + \beta_5}, \eqno(6.2)$$
$$y(x,z)=-{\b_1 xz+\b_5 x+\b_4z+\b_7\over\b_3x+\b_2z+\b_6},\eqno(6.3)$$
and
$$z(x,y)=- {\b_3xy+\b_5x+\b_6y+\b_7\over\b_1x+\b_2y+\b_4}.\eqno(6.4)$$
The surface $z(x,y)$ consists of a pair of disconnected hyperbolic
sheets as shown in Figure 2.  The sheets are separated by the plane
$\b_1x+\b_2y+\b_4=0$, where the denominator of (6.4) vanishes.  If
these sheets pass through the
region of integration of (5.2), which is a rectangular box with opposite
corners $(0,0,0)$ and $( \Delta x, \Delta y, \Delta z)$, then the
$\delta$ function will have support in that region and $S$ may
contribute to $E$.

The hyperbolic sheets have a simple physical interpretation.  For
purposes of clarity, we discuss the case $S^{\vt}_{ijkl}$; the other
three integrals in (5.4) have similar interpretations.  The $z$
coordinate in $S$ parameterizes the world-line followed by the $k$th
kink on the string loop.  This kink moves along a straight, null
world-line.  The $x$, $y$ integrations are over a (diamond-shaped)
planar patch of the world-tube, swept out by a linear segment of the
string loop.  Note that these patches are always time-like.  The edges
of this planar patch are bounded by the straight, null world-lines of
$i$th and $j$th kinks.  The future and past light-cones of any point on
the $k$th kink's world line (i.e., fixed $z$) may intersect the planar
patch bounded by the $i$th and $j$th kinks.  Even if they fail to
intersect this patch, they will intersect the infinite 2-plane passing
through the patch, which is parameterized by $x$ and $y$.  The
intersection will trace out a hyperbola on the plane which may or may
not intersect the actual $x$, $y$ integration region.  The hyperbola is
given (with $z$ fixed) by (6.3).   This hyperbola corresponds to the
intersection of a $z=$\ constant plane with the hyperbolic sheets of
Figure 2.  Because one branch of this hyperbola lies on the future
light-cone and the other lies on the past light-cone, we refer to these
as the {\it future branch} and the {\it past branch}.  The two branches
are disconnected except in the case where the plane passes through the
origin of the light cone.  In this case they touch at a single point -
the apex of the light-cone.  As the apex moves along the world-line of
the kink (i.e., $z$ increases) the light-cone sweeps out a region on
the x-y plane.  This can be seen by taking successive $z=$\ constant
cross sections of the hyperbolic sheets, where the constant ranges from
0 to $\Delta z$.  The region swept out in the x-y plane will be bounded
by the hyperbolae $y(x,0)$ and $y(x,\Delta z)$.  Note that if we
restrict attention to just the future branches or just the past
branches, then the $y(x,\Delta z)$ hyperbola always lies above (and to
the right of) the $y(x,0)$ hyperbola in the x-y plane.  Therefore, we
refer to these as the ``Top" and ``Bottom" hyperbolae, as shown in
Figure 3.  The hyperbolae shown in Figure 3 will intersect the $x$, $y$
integration region only if the hyperbolic sheets pass inside the
integration box shown in Figure 2.

There are two useful sets of conditions which can be checked
immediately to see if $S$ vanishes.  These test whether the hyperbolic
sheets $z(x,y)$ pass through the integration box.  If the sheets do not
pass through the box, then the $\delta$ function in (5.2) has no
support, and $S$ vanishes.  One of the conditions applies for $s=+1$.
In this case, the $\theta$ function in (5.2) restricts the integration
to be over only the hyperbolic sheet swept out by the future
light-cone.  The other condition applies for $s=-1$. In this case, the
$\theta$ function in (5.2) restricts the integration to be over only
the hyperbolic sheet swept out by the past light-cone.   If $s=+1$,
then
$$\eqalign{
S=0\iff\cases{
\tau+\Delta x+\Delta y\leq 0 ,\rm{or} &\cr&\cr
\b_3\Delta x \Delta y+\b_5\Delta x+\b_6 \Delta y+\b_7\geq 0 , \rm{or}
&\cr&\cr\Delta z\leq \tau \quad\rm{and}\quad \b_4\Delta z+\b_7\leq 0.
\cr}\cr}\eqno(6.5)$$
Similarly, if $s=-1$, then
$$\eqalign{
S=0\iff\cases{
\tau+\Delta x\leq 0 , \rm{or} &\cr&\cr
\b_4\Delta z+\b_7\geq 0 , \rm{or}&\cr&\cr
\Delta x+\Delta y\leq\tau \quad\rm{and}\quad\b_3\Delta x\Delta y
+\b_5\Delta x+\b_6\Delta y+\b_7\leq 0.\cr}\cr}\eqno(6.6)$$
In practice, these conditions are frequently satisfied,
so their implementation saves large amounts of computing time.

It is straightforward to do the integral over $z$ in (5.2) to
eliminate the $\delta$ function in $S$.  We make use of the
standard formula
$$\int \delta(f(z)) g(z) dz = \sum_p g(z_p)/
\big |{\partial f\over \partial z}(z_p)\big |\eqno(6.7)$$
where the sum is taken over all the roots $z_p$ of $f(z)$ that lie
within the range of $z-$integration.  Because
$${\partial f\over \partial z}=\b_1x+\b_2y+\b_4,\eqno(6.8)$$
the integration of (5.2) over $z$ yields
$$\eqalign{S(\Delta x,\Delta y,\Delta z,& \tau,s,C,\beta_1,\cdots,\beta_7)
\cr &=\int\limits_0^{\Delta x}dx
\int\limits_{0}^{\Delta y}dy\>
{\theta(z(x,y))
\theta(\Delta z - z(x,y))
\theta(\tau+s(x+y-z(x,y))) \over |\b_1x+\b_2y+\b_4|}.\cr}
\eqno(6.9)$$
The first two $\theta$ functions in this equation arise from (6.7):
the only roots included in the sum over $p$ are those lying within the
range $z\in [0,\Delta z]$ of $z$-integration.

The integral (6.9) has a simple physical interpretation which is
directly connected to the physical interpretation of (5.2) already
given following equation (6.4).  The $x $ and $ y$ integrations are
over a planar patch on the string loop's world-tube, as in (5.2).  The
first two $\theta$ functions in (6.9) only have support between the
hyperbolic curves $y(x,0)$ and $y(x,\Delta z)$ (i.e., the Bottom and
Top curves) shown in Figure 3.  The curve $y(x,0)$ describes the
intersection of the hyperbolic sheet shown in Figure 2 with the bottom
of the integration box.  The curve $y(x,\Delta z)$ describes the
intersection of the hyperbolic sheet with the top of the integration
box.  Thus, the first two $\theta$ functions in (6.9) will only have
support if the hyperbolic sheets pass through the box of integration in
(5.2).  The third $\theta$ function in (6.9) effectively restricts the
integration to be over the future branches ($s=+1$) or the past
branches ($s=-1$) of the hyperbolae.

When evaluating (6.9), there are five different fundamental types of
integrals that can arise depending upon the relative positions of the
two hyperbolic curves $y(x,0)$ and $y(x,\Delta z)$ and the rectangular
region in the x-y plane bounded by the opposite corners $(0,0)$ and
$(\Delta x,\Delta y)$.  Each of the five possibilities are shown (using
the future branches of the hyperbolae) in Figure 4.  Each type of
integral may also occur with the past branches of the hyperbolae.  The
first type occurs when both hyperbolae (Top and Bottom) pass through
the planar patch.  The second type occurs when the entire planar patch
lies between the two hyperbolae.  A third possible type occurs when
neither hyperbola passes though the planar patch and the planar patch
does not lie between the two hyperbolae.  For this type of integral,
(6.9) has no support and vanishes.  The last two types arise when one
of the hyperbolae (Top or Bottom) passes through the planar patch but
the other does not.

Remarkably, each of the fundamental integral types can be done
analytically in closed form.  To assist in this process, the $x$
integration in (6.9) must be divided into consecutive ranges; the type
of integral in each range is different.  A systematic method for
determining these integration ranges in $x$ will be given next.

The dividing points (in $x$) between the successive ranges are
determined by the four points at which the bottom and top hyperbolae
$y(x,0)$ and $y(x,\Delta z)$ intersect the lines $y=0$ and $y=\Delta
y$.  We denote the $x$ coordinates of the four intersection points by
$\phi_1, \dots , \phi_4$.  The type of integral being done in (6.9)
will change at each of the intersection points which is within the
boundaries of the $x$ integration, i.e. $0<x<\Delta x$.  The boundaries
of the $x$ integration will be labeled $\phi_0$ and $\phi_5$.  The $x$
coordinates $\phi_1, \dots , \phi_4$ are given by;
$$\eqalign{
    \phi_1 = & x(0,\Delta z)=- {\b_4 \Delta z + \b_7 \over
        \b_1\Delta z+\b_5}\quad\rm{if}\quad C\ne 1\quad\rm{else}\quad
\phi_1=0,\cr
    \phi_2 = & x(\Delta y,\Delta z)=- {\Delta z (\b_2 \Delta y + \b_4)+
       \b_6 \Delta y + \b_7 \over \b_3 \Delta y +\b_1 \Delta z +
	  \b_5}\quad\rm{if}\quad C\ne 2\quad\rm{else}\quad \phi_2=0,\cr
    \phi_3 = & x(\Delta y,0)=- {\b_6 \Delta y + \b_7 \over \b_3 \Delta y
	+ \b_5}\quad\rm{if}\quad C\ne 3\quad\rm{else}\quad \phi_3=0, \cr
    \phi_4 = & x(0,0)=-{\b_7 \over \b_5}\quad\rm{if}\quad C\ne 4\quad\rm{else}
	\quad \phi_4=0. \cr} \eqno(6.10)$$
The values of $C$ are checked because there are four special cases.  In
these special cases the formula given for one of $\phi_1, \dots ,
\phi_4$ is indeterminate because the numerator and denominator in
(6.10) vanish.

The conditions for the four special cases are given in the definition
(5.3) of $C$.  Physically, these special cases arise when one end of
the kink's world-line, parameterized by $z$, touches one of the four
corners of the diamond-shaped patch of the world-sheet defined by the
$x$ and $y$ integrations.  The intersections of the future and past
light cones of a point on the kink's world-line with the plane defined
by the diamond-shaped patch are usually hyperbolae.  However, when the
point on the kink's world-line is also a corner of the diamond-shaped
region, then the plane passes through the apex of the light cone and
the hyperbola degenerates into a pair of straight lines.  These
straight lines will lie along the two edges of the planar patch which
are joined at the corner where the kink's world-line touches.  For each
of the four special cases, one of the formula for $\phi$ in (6.10)
would become indeterminate.  Consider, for instance, the case where the
end of the kink's world-line (ie.  $z=\Delta z$) touches the lower left
corner of the integration region (ie. $x=y=0$).  At this point, we have
$x^\alpha(u,v)=x^\alpha(\ut,\vt)$.  In this case the future top curve
lies along the left and lower sides of the rectangular integration
region.  One can verify that in this case, both the numerator and
denominator in the equation for $\phi_1$ vanish.  It is $\phi_1$ that
becomes indeterminate in this case because $\phi_1$ is the $x$
coordinate of the intersection of the top curve and the line $y=0$,
which does not have a unique solution in this case.  The other three
special cases are similar.  In each case a different $\phi$ would
become indeterminate if the value of $C$ were not checked.

All four special cases are dealt with in the same manner.  The purpose
of the $\phi$'s is to locate the $x$ coordinates where the type of
integral being done changes.  However, since intersection curves which
lie along the edges of the integration region never cause the type of
integration being done to change, it is sufficient to simply set the
corresponding $\phi_i$ to zero (or to any value outside the range
($\phi_0,\phi_5$)).

The support of the $x$ integration in (6.9) may be less than the range
$0<x<\Delta x$ because of the third $\theta$ function.  Thus one does
not always have $\phi_0 =0$ and $\phi_5 =\Delta x$.  Because the
integrations in (6.9) have support only between the future branches
($s=+1$) or the past branches ($s=-1$) of the hyperbolae, it is
convenient to define $\phi_0$ and $\phi_5$ in a more general way.
First, we define $\phi_B$ to be the vertical asymptote to the bottom
hyperbola when $s=+1$ and zero otherwise, and $\phi_T$ to be the
vertical asymptote to the top hyperbola when $s=-1$ and zero otherwise,
$$\eqalign{
	\phi_B= &\theta(s) x(y \to \infty,0)
                   =-\theta(s) {\beta_6 \over \beta_3}\cr
	\phi_T= &\theta(-s) x(y \to \infty,\Delta z)
                   =-\theta(-s){\beta_2 \Delta z + \beta_6 \over \beta_3}.\cr}
\eqno(6.11)$$
The boundaries of the $x$ integration are then defined to be
$$\eqalign{
	\phi_0 = & \max(0,\phi_B)\cr
	\phi_5 = & \min(\Delta x,\phi_T+\theta(s)\Delta x ).\cr}\eqno(6.12)$$
This definition of $\phi_0$ and $\phi_5$ eliminates regions of
integration which only contain the past branches of the hyperbolae when
$s=+1$ and regions which only contain the future branches of the
hyperbolae when $s=-1$.

One may now express $S$ as a sum of integrals over the successive ranges
of the $x$-integration.
The set of points $\{\phi_0 \dots \phi_5\}$ partition the
$x$-integration into at most five ranges.  Let
$\{x_0,x_1,x_2,x_3,x_4,x_5\}$ be the increasing sorted set of $\phi$'s,
$$ \{x_0,x_1,x_2,x_3,x_4,x_5\} =
{\rm sort}(\{\phi_0,\phi_1,\phi_2,\phi_3,\phi_4,\phi_5 \}), \eqno(6.13) $$
sorted so that $x_i\leq x_{i+1}$.
If we define the mid-point between two successive $x$'s as
$${\bar x}_n = (x_n+x_{n+1})/2, \eqno(6.14)$$
then we may rewrite the $S$ integral (6.9) as a sum of integrals over
each successive $x$ range;
$$S(\Delta x,\Delta y,\Delta z, \tau,s,C,\beta_1,\cdots,\beta_7) =
\sum_{n=0}^4 \theta({\bar x}_n - \phi_0) \theta(\phi_5 - {\bar x}_n)
T(x_n,x_{n+1},\Delta y,\Delta z, \tau,s,\beta_1,\cdots,\beta_7).
\eqno(6.15)$$
The two $\theta$ functions ensure that only the ranges of $x$ between
$\phi_0$ and $\phi_5$ may contribute.  The function $T$ is defined to be
$$T(x_l,x_u,\Delta y,\Delta z, \tau,s,\beta_1,\cdots,\beta_7)=
\int\limits_{x_l}^{x_u}dx
\int\limits_{0}^{\Delta y}dy\>
{\theta(z(x,y))
\theta(\Delta z - z(x,y))
\theta(\tau+s(x+y-z(x,y))) \over |\b_1x+\b_2y+\b_4|}.\eqno(6.16)$$
The reason that we have defined the function $T$ is that the $x$
integration being done in (6.16) is over a region which contains only
one of the five possible fundamental types of integrals discussed in
the paragraphs preceding equation (6.10) and shown in Figure 4.

What remains is to find the analytic form of (6.16) for each of the
five possible types of integrals that can arise.  Recall that the
different types of integrals arise, as shown in Figure 4, from the
different possible relative positions of the Top and Bottom hyperbolae
compared to the $(x,y)$ region of integration.  If the region of
integration is between the two hyperbolae, then the first two $\theta$
functions in (6.16) will have support over the entire region, and the
limits of the $y$ integration will run from $0$ to $\Delta y$.  If both
hyperbolae pass through the region of integration, then the first two
$\theta$ functions in (6.16) restrict the $y$ limits of integration to
run from $y(x,0)$ to $y(x,\Delta z)$.  The last two types of integrals
which give non-zero contributions are when one of the hyperbolae passes
through the region of integration but the other does not.  These types
of integrals will have $y$ limits of integration which run from $0$ to
$y(x,\Delta z)$ or from $y(x,0)$ to $\Delta y$ depending on whether it
is the Top or the Bottom hyperbola that passes through the region of
integration.  For each type of integral, the third $\theta$ function in
(6.16) restricts the integration region to be between either the future
($s=+1$) or the past ($s=-1$) branches of the hyperbolae.  If we make
the definitions,
$$\eqalign{{\bar x} =& (x_l+x_u)/2, \cr
y_b=& s \> y(\bar x,\theta(-s) \Delta z),\cr
y_t=& s \> y(\bar x,\theta(s) \Delta z),\cr
y_l=& -\theta(-s) \Delta y,\cr
y_u=& \theta(s) \Delta y,\cr
\delta =& \epsilon( \b_1 \bar x+ s \beta_2 y_b +\b_4),\cr} \eqno(6.17)$$
then the integrals in (6.16) can be carried out for each of the five
different cases.  One obtains
$$\eqalign{
T(&x_l,x_u,\Delta y,\Delta z, \tau,s,\beta_1,\cdots,\beta_7)\cr
=&\cases{
\displaystyle \int_{x_l}^{x_u} dx \>
\int_{0}^{\Delta y} dy\> {1 \over |\b_1x+\b_2y+\b_4|} =
   \cr
\displaystyle {\delta \over \beta_2} \biggl(
L(x_l, \>x_u, \>\beta_1, \>\beta_4+ \beta_2 \Delta y ) -
L(x_l, \>x_u, \>\beta_1, \>\beta_4) \biggr)
  & for $y_t \le y_b \le y_l$, or
\cr
   & for $y_b \le y_l$ and $y_u \le  y_t$\cr\cr\cr
\displaystyle  \int_{x_l}^{x_u} dx \>
\int_{\theta(-s) y(x,0)}^{\theta(s) y(x,\Delta z)+\theta(-s) \Delta y} dy\>
{1 \over |\b_1x+\b_2y+\b_4|} =\cr
\displaystyle { s \delta \over \beta_2} \biggl( Q(x_l, \>x_u,
\>\beta_1 \beta_3, \>
    \beta_1 \beta_6+\beta_3 \beta_4 - \beta_2 \beta_5, \>
    \beta_4 \beta_6 - \beta_2 \beta_7)  \cr
\> \quad - L(x_l, \>x_u, \>\beta_3, \>  \beta_6 +
\beta_2 \theta(s) \Delta z) -
L(x_l, \>x_u, \>\beta_1, \>\beta_4+\beta_2 \theta(-s) \Delta y)\biggr)
    & for $y_b \le y_l < y_t < y_u$\cr\cr\cr
\displaystyle  \int_{x_l}^{x_u} dx \>
\int_{y(x,0)}^{y(x,\Delta z)} dy\> {1 \over |\b_1x+\b_2y+\b_4|} =\cr
\displaystyle {\delta \over \beta_2} \biggl( L(x_l, \>x_u,
\>\beta_3, \>\beta_6) -
L(x_l, \>x_u, \>\beta_3, \> \beta_6 + \beta_2 \Delta z )\biggr)
    & for $y_l < y_b < y_t <y_u $\cr\cr\cr\cr
\displaystyle  \int_{x_l}^{x_u} dx \>
\int_{\theta(s) y(x,0)}^{\theta(-s) y(x,\Delta z)+\theta(s) \Delta y} dy\>
{1 \over |\b_1x+\b_2y+\b_4|} =\cr
\displaystyle { s \delta \over \beta_2} \biggl(
L(x_l, \>x_u, \>\beta_3, \>\beta_6+\beta_2 \theta(-s) \Delta z) +
L(x_l, \>x_u, \>\beta_1, \> \beta_4+\beta_2 \theta(s) \Delta y)
  \cr
\quad - Q(x_l, \>x_u, \>\beta_1 \beta_3, \>\beta_1 \beta_6+\beta_3 \beta_4
  - \beta_2 \beta_5, \>\beta_4 \beta_6 - \beta_2 \beta_7) \biggr)
  & for $y_l < y_b < y_u \le y_t$, or \cr
  & for $y_l < y_b < y_u$ and $y_t \le  y_b$\cr\cr\cr\cr
\displaystyle \qquad \qquad \qquad \qquad \qquad \qquad 0
    & for $y_u \le  y_b$, or\cr
    & for $y_b < y_t \le y_l$.\cr}\cr}
\eqno(6.18)
$$
Here, the functions $L$ and $Q$ are the ``linear" and ``quadratic"
integrals defined by
$$\eqalign{L{\bigl(}& x_1,x_2,c_1,c_0 \bigr) =
\int\limits_{x_1}^{x_2}dx\>\ln |c_1x+c_0| \cr
&= \cases{
\bigg[\big(x+{c_0\over c_1}\big)\ln |c_1x+c_0|-x\bigg]_{x_1}^{x_2}
& for $c_1 \ne 0$ \cr\cr\cr
(x_2 - x_1) \ln  | c_0 | & for $c_1 = 0 $\cr}\cr}\eqno(6.19)$$
and
$$\eqalign{
Q\bigl( & x_1,x_2,c_2,c_1,c_0 \bigr)  =
\int\limits_{x_1}^{x_2}dx\>\ln |c_2x^2+c_1x+c_0| \cr
&=\cases{
L{\bigl(}x_1,x_2,c_1,c_0 \bigr)
& for $c_2=0$ \cr\cr\cr
L{\bigl(}x_1,x_2,1,{c_1+\sqrt{c_1^2 - 4 c_0 c_2} \over 2 c_2}\bigr) +
L{\bigl(}x_1,x_2,1,{c_1-\sqrt{c_1^2 - 4 c_0 c_2} \over 2 c_2}\bigr)+ & \cr
\quad  (x_2-x_1) \ln | c_2| & for
$0 \le c_1^2 - 4 c_0 c_2 $\cr\cr\cr
\biggl[ x \ln (x^2 + { 4 c_0 c_2 - c_1^2 \over 4 c_2^2}) -  2 x
  + 2 \sqrt{{ 4 c_0 c_2 - c_1^2 \over 4 c_2^2}}
  \arctan(x \sqrt{{4 c_2^2 \over 4 c_0 c_2 - c_1^2}} )
  \biggr]_{x_1+{c_1 \over 2 c_2}} ^{x_2+{c_1 \over 2 c_2}}+ & \cr
 \quad L{\bigl(}x_1+{c_1 \over 2 c_2}, x_2+{c_1 \over 2 c_2},0,c_2\bigr)
&  for $c_1^2 - 4 c_0 c_2 <0.$ \cr}\cr}\eqno(6.20)$$
Thus $T$, and hence $S$ has been evaluated analytically for all
possible cases.  Using the results of this section, one can carry out
the summations in (5.1) to arrive at a final value for $\gamma$; the
power radiated in gravitational waves by a string loop.

\vskip 0.3in
\centerline{Section 7: TESTING THE FORMULA FOR $\gamma$ AGAINST PREVIOUS
RESULTS}
\vskip 0.05in

In this section we compare the $\gamma$ values given by our formulae to
previously published values for a large number of loop trajectories.
The formulae obtained in sections 5 and 6 were directly implemented by
computer code.  In some cases we find disagreement between our results
and those previously published.  There are, in fact, several cases
where conflicting results have been published for the same loop
trajectories.  In the cases where a disagreement was found, we have
identified the errors made in the published work which led to the
incorrect results.  In these cases our formulae give the correct values
of $\gamma$.  We are confident that they are correct because, in every
case, we have shown our results to be consistent with those given by
other independent methods.  The other methods used to confirm our
results were the FFT method of Allen and Shellard [14] and/or a
corrected implementation of the numerical method used by the original
author(s).

While our formulae handle piecewise linear loops exactly, most of the
previous work in this area has considered smooth cosmic string
trajectories (typically providing analytic expressions for the a- and
b-loop).  To compare the results of our formulae to the published
values of $\gamma$ for these smooth loops, we calculate $\gamma$ for
piecewise linear loop trajectories of approximately the same shape.  If
the number of segments used ($N_a$ and $N_b$) is reasonably large, then
the piecewise linear loop trajectory and the smooth loop trajectory
will be very similar in shape, and we expect that the values of
$\gamma$ for the two trajectories will be very close.  An example of
how we generate specific piecewise linear trajectories is given later
in this section.  The rate at which $\gamma$ converges as the number of
linear segments is increased is also discussed.

Prior to this work, the only fully analytic closed form solution for
$\gamma$ for any string loop trajectory was given by Garfinkle and
Vachaspati [11].  They considered the piecewise linear loops defined by
a- and b-loops which consisted of just two linear segments each; i.e.
$N_a=N_b=2$.  This defines a family of loop trajectories which depend on
a single parameter $\phi$, the angle  between the a- and b-loops.   As a
function of $\phi$, $\gamma$ is given by
$$\gamma(\phi)={32\over \sin ^2\phi}\biggl[(1+\cos\phi)\ln\bigl({2\over
1+\cos\phi}\bigr)+(1-\cos\phi)\ln\bigl({2\over 1-\cos\phi}\bigr)\biggr].
\eqno(7.1)$$
When calculating $\gamma$ with our formulae, we used a- and b-loops with
three segments (where the length of the third segment was much smaller
than the other two).  This was necessary in order to prevent a singularity;
exactly parallel segments cause the determinant in (4.14) to vanish.
Equation (7.1) is plotted in Figure 5 (solid line) along with the $\gamma$
values (dots) given by our code for a number of loops with different values
of $\phi$.  Since these are piecewise linear loops, we expect our
results to be highly accurate.  Indeed, the points plotted in Figure 5
showing our results had to be enlarged in order to distinguish them
from the plot of (7.1).  Thus our method completely confirms the results
of Garfinkle and Vachaspati.

The next set of loop trajectories which we use to test our formulae is
a three-parameter family of trajectories first examined by Burden [10].
The three parameters are $L$, $M$ and $\phi$, where $L$ and $M$ are
positive integers and $\phi$ is an angle in the range $[0,\pi]$.
The Burden trajectories are defined by the a- and b-loops:
$$\eqalign{\vec a(u)&={L^{-1}\over 2\pi}[\cos(2\pi Lu)\hat z+\sin(2\pi Lu)
(\cos\phi\hat x +\sin\phi \hat y)],\cr
\vec b(v)&={M^{-1}\over 2\pi}[\cos(2\pi Mv)\hat z-\sin(2\pi Mv)\hat x].\cr}
\eqno(7.2)$$
The b-loop winds $M$ times around a circle in the x-z plane. The a-loop
winds $L$ times around a circle whose plane is at an angle $\phi$ with
respect to the x-z plane.  The Burden string loops are non-intersecting
cuspy loops in the case $M=1$, $L>1$ and $\phi$ not equal to 0 or
$\pi$.  Burden calculated values of $\gamma$ for loops with $M=1$ and
$L=1,2,3,5,15$ for several values of $\phi$.  The values of $\gamma$
for loops with $M=1$ and $L=3,5$ were also calculated by Quashnock and
Spergel [13].  Using our formula, we calculated the values of $\gamma$
for a large number of loops, each of which is a piecewise linear
approximation to a Burden loop.  In addition, we calculated a number of
$\gamma$ values using the FFT method of Allen and Shellard [14].  Our
results for loops with $M=1$ and $L=3,5$ are shown in Figure 6 along
with the results of Burden, Quashnock and Spergel, and the FFT method.
We find excellent agreement among all four sets of results.  This also
shows that piecewise linear loops with fairly small numbers of segments
($N_a=16L$ and $N_b=16M=16$) can provide excellent approximations to
smooth loop trajectories and provides further evidence that our
formulae are correct.

Values of $\gamma$ for the Burden loops with $L=M=1$ have been
published by Burden [10], Vachaspati and Vilenkin [9], and Durrer [8].
These results, along with the results of the FFT method and our new
method are shown in Figure 7.  There is excellent agreement between
four of the sets of results.  However, Durrer's results for these
trajectories do not agree well with the others.

To understand why Durrer's results do not agree with the others, we
recalculated $\gamma$ for these trajectories using the same numerical
method used by Burden, Vachaspati and Vilenkin, and Durrer.  This
method requires one to calculate the average power radiated by a string
loop using the formula previously given in (3.7)
$$P=\sum_{n=1}^\infty\int d\Omega {dP_n\over d\Omega},
\eqno(7.3)$$
where $P_n$ is the average power radiated at frequency $\omega_n=4\pi
n$ and the integration is over the 2-sphere.  The details of the
calculation of $P_n$ can be found in references [8,9,10].  (Note however
the following typographical errors in reference [8]. The term
$J_{l+1}(-l\sin\phi)$ in (A.6) should be $J_{l+1}(-l\sin\theta)$ and $y$
should be replaced by $-y$ in (A.12) and (A.13).) Because of the infinite
sum appearing in (7.3), one must stop calculating the $P_n$ numerically
at some value of $n$, and then estimate the contribution to the sum from
larger values of $n$.  Since the sum may be slowly convergent, this
``tail" may give a significant contribution even when the individual
$P_n$ are very small.  For the $L=M=1$ Burden loops with $\phi\ne 0$ or
$\pi$, the tail can estimated with good accuracy because the $P_n$ fall
off as a power law $n^{-4/3}$ for large $n$.  Durrer's results are
incorrect precisely because the contribution of the tail was not included
at all.  In Figure 8 we show Durrer's original results and our calculation
of the sum of the first 50 terms of (7.3).  Note the agreement between
these values.  We also show the results of the first 50 terms plus an
estimate of the tail of the sum along with the results from our code.
It is clear that when the tail is included, Durrer's results then agree
with the results found by all other investigators.  Thus, again we find
that our method is in agreement with the results of previous authors.

Before continuing to compare the results of our new formulae to those
in the published literature, we take a moment to discuss how piecewise
linear loop trajectories are constructed to approximate smooth loop
trajectories.  We illustrate the procedure by explaining how the
piecewise a- and b-loops were constructed to approximate the $L=M=1$
Burden loops considered above.  The piecewise linear a-loop was
constructed by dividing the interval $u\in [0, 1]$ into $16$ equal
segments.  This defines $16$ values of the coordinate $u$.  These
values were then perturbed by small random amounts so that pairs of
segments on the a-loop would not end up exactly parallel.  (This is
necessary to prevent the determinant in (4.14) from vanishing.)  The
perturbed values of $u$ were then used in the first equation of (7.2)
to yield the coordinates of the $N_a=16$ kinks which define the
a-loop.  Finally, the entire a-loop was translated in 3-space so that
the kink with parameter $u=0$ was positioned at the origin.  The
piecewise linear b-loops were constructed in a similar manner.  Each
b-loop was constructed to have $N_b=16$ linear segments.  In all cases,
our values of $\gamma$ were within 8 percent of previously calculated
results with an average difference of less than 3.5 percent.  Thus, for
the purposes of calculating $\gamma$, a single wind around a smooth
circular path  is approximated extremely well by a set of only 16
linear segments.

We now examine how the $\gamma$ values found for the piecewise linear
approximation to smooth loop trajectories depends on the number of
segments used.  To determine this dependence, we constructed several
piecewise linear approximations to the $L=M=1$ Burden loops using the
procedure given above, each with different values of $N_a$ and $N_b$.
The results of four such tests are shown in Figure 9.  One can see that
the values of $\gamma$ converge quickly as $N_a$ and $N_b$ increase.
It is only for loops with values of $\phi$ near 0 and 180 degrees
(where $\gamma$ diverges) that a large number of segments are needed to
obtain good accuracy.  The relative errors in four approximations
compared to the most accurate approximation (curve D in Figure 9) are
given in Figure 10.  The errors decrease rapidly as the number of
segments increases.  These errors are small and are mainly due to the
loops which have $\phi$ close to 0 or 180 degrees.  We obtained similar
results for the $L=3, M=1$ and $L=5, M=1$ Burden loops.  Further
discussion of how the accuracy of $\gamma$ in the piecewise linear
approximation of a smooth loop depends on the number of segments
$N_a, N_b$ is postponed until the end of this section.

We now continue to compare the results of our formulae to those in the
published literature.  The next set of loop trajectories with which we
compare our results is a two-parameter family of loops first studied by
Vachaspati and Vilenkin [9].  The a- and b-loops which define these
trajectories are given by
$$\eqalign{\vec a(u)=&{1\over 2\pi}[\sin(2\pi u)\hat x
-\cos(2\pi u)(\cos\phi\hat y
+\sin\phi\hat z)]\cr \vec b(v)=&{1\over 2\pi}\bigl[\bigl({\alpha\over 3}
\sin(6 \pi v)-(1-\alpha)\sin(2\pi v)\bigr)\hat x \cr
&\quad\quad-\bigl({\alpha\over 3}\cos(6 \pi v)+
(1-\alpha)\cos(2\pi v)\bigr)\hat y
\cr&\quad\quad\quad\quad\quad-(\alpha(1-\alpha))^{1/2}\sin(4\pi v)
\hat z\bigr]\cr}
\eqno(7.4)$$
where $\alpha$ and $\phi$ are constant parameters, $0\le\alpha\le 1$
and $-\pi <\phi\le\pi$.  Note that when $\alpha=0$, these trajectories
are equivalent to the $L=M=1$ Burden loops studied above.  These loops
have also been studied by Durrer [8].  The results found by Vachaspati
and Vilenkin, and Durrer are shown in Figure 11 along with the results
of the FFT method and the results of our new method for the case
$\alpha=0.5$.  We find that only the FFT method and our new method are
in good agreement.  The $\gamma$ values given by Vachaspati and
Vilenkin, and Durrer appear to be too small.  In fact, their results
are lower than the sum of the first 300 $P_n$ found by the FFT method
(see Figure 11).  We take the sum of the first 300 $P_n$ to be a lower
bound for $\gamma$ since continuing the sum to larger $n$ or adding an
estimate of the tail (or both) will only increase the value found for
$\gamma$.

There are two possible explanations for the incorrect results given by
Vachaspati and Vilenkin.  The first possibility is that they
incorrectly estimated the tail contribution to the sum in (7.3).
Vachaspati and Vilenkin claim that the sum in (7.3) is rapidly
convergent, with $P_n\propto n^{-3}$ for large $n$.  However, we have
found that the sum is actually much less convergent.  For example, the
power spectrum for the trajectory with $\alpha =0.5$ and $\phi=\pi/2$
is shown in Figure 12.  In this case, $P_n\propto n^{-1.25}$ for $n$ in
the range $100<n<300$.  By overestimating the convergence of the sum in
(7.3), one seriously underestimates the contribution due to the tail of
the sum.  The other possible explanation is that the results reported
in [9] are actually for a different set of loops than those defined by
(7.4).  We consider this a possibility not only because the reported
convergence of the sum (7.3) does not agree with our findings, but
also  because Vachaspati and Vilenkin include a drawing (Figure 4 of
reference [9]) of the loop's shape at two different times during its
oscillation.  However, these shapes do not agree with the shapes given
by (7.4).  We have confirmed that the loops shown in Figure 4 of
reference [9] are not the same as the loops given by (7.4), however we
have been unable to resolve whether the values of $\gamma$ reported in
reference [9] correspond to the loops defined by (7.4) or to those
shown in the figure [17].

We have calculated values of $\gamma$ for the loop trajectories (7.4)
using our new formulae for several other values of the parameter
$\alpha$.  When $\alpha=0$, the loop trajectory is equivalent to the
$L=M=1$ Burden loops.  Thus, for small $\alpha$, the loop trajectories
(7.4) should be similar to those given by (7.2).  In Figure 13 we show
our results for $\alpha=0.01$ (solid line).  This is compared to
results using the traditional numerical method (crosses) and the
results for $\alpha=0$ (dashed line).  The results of our new method
agree well with those of the traditional numerical method.

The final string loop trajectories with which we compare our formulae
were first given by Garfinkle and Vachaspati (equation (2.9) of
reference [11]).  The a- and b-loops for these trajectories are
composed of two smooth circular arcs joined by a pair of straight
segments.  The analytic expressions for the a- and b-loops are
$$\eqalign{\vec a(u)=&{1\over 2\pi q}
\bigl[\sin(\delta (u)+2\pi qu)\hat x-\cos(\delta (u)+2\pi qu)
(\cos\phi\hat y+\sin\phi\hat z)\bigr]\cr\vec b(v)=&{1\over 2\pi p}
\bigl[\sin(\beta (v)-2\pi pv)\hat x-\cos(\beta (v)-2\pi pv)\hat y\bigr].
\cr}\eqno(7.5)$$
Here, $p$ and $q$ are constants in the range $[0,1]$, $\phi$ is the
angle between the two loops, and $\beta$ and $\delta$ are defined by
$$\beta(v)=(1-p)\pi[-2v]\quad\rm{and}\quad\delta(u)=(1-q)({\pi\over
2}+\pi [2u]).\eqno(7.6)$$ In (7.6), $[x]$ is the greatest integer less
than or equal to $x$.  Our results for trajectories with
$(p,q)=$(0.6,0.4), (0.4,0.8) and (0.9,0.9) are shown in Figure 14.  The
results of the FFT method for several trajectories with
$(p,q)=$(0.9,0.9) are also shown in Figure 14.  Again, we find good
agreement between the two methods.  Garfinkle and Vachaspati do not
give specific values but claim that the trajectories given in (7.6)
have $\gamma$ values around 100.  This is consistent with our results.
Durrer [8] has also given values of $\gamma$ for some of these
trajectories.  For the three trajectories with parameters
$(p,q)=$(0.6,0.4), (0.4,0.8) and (0.9,0.9) and with $\phi=\pi/2$,
Durrer reports $\gamma$ values of 19, 26 and 42 respectively.  However,
because of the errors (explained above) in other numerical results
presented by Durrer, we do not have confidence in these values of
$\gamma$.  Durrer's results appear to be too low, which would be
consistent with leaving off the contribution from the tail of the sum
in (7.3).  The agreement between the FFT method and our new method
again gives us confidence that our formulae are correct.

We now return to the question of how accurately the $\gamma$
values from piecewise linear loop trajectories approximate the $\gamma$
values from smooth loop trajectories.  In particular, we would like to
know how the difference between the $\gamma$ value of a piecewise linear
loop with $N=N_a+N_b$ total segments ($\gamma_N$) and the $\gamma$
value of the smooth loop that it approximates ($\gamma_{\infty}$) falls
off as a function of $N$.  Unfortunately, we do not know how the
difference $\Delta(N)\equiv|\gamma_\infty-\gamma_N|$ depends on $N$ in
the general case.  However, numerical estimates may be made for
individual loops with the hope that the results will hold in general.
In addition, there is at least one case which has been investigated
where simple analytic formulas exist for both $\gamma_\infty$ and
$\gamma_N$.

A detailed numerical investigation of $\Delta(N)$ has been carried out
for two Burden loop trajectories (7.2) with $L=M=1$.  The numerical
values of $\gamma_N$ have been computed over a wide range in $N$ for
both loops.  The value of $\phi$ was arbitrarily chosen to be
$39^\circ$ for the first loop, and $111^\circ$ for the second loop (see
Figure 15).  We find that for large $N$, both sets of results are well
approximated by functions of the form $A+B N^{-1}$, where $A$ and $B$
are constants that depend only upon $\phi$.  For the first set of
results ($\phi=39^\circ$), we find that $\gamma_N\approx 52.01+181.64
N^{-1}$ for $60\leq N \leq 256$.  By taking $\gamma_\infty=52.01$, we
can find a numerical estimate of $\Delta(N)$.  A similar analysis has
been done for the second set of loops, where $\gamma_N\approx
64.49+97.13 N^{-1}$.  Figure 16 shows a Log-Log plot of $\Delta(N)$ for
both sets of loops.  By examining the slopes of the curves in Figure
16, we find that in both cases, $\Delta(N)$ falls off like $N^{-1}$ for
large $N$.

In addition to the numerical investigations of $\Delta(N)$ given above,
there is one case where $\Delta(N)$ has been calculated analytically.
In a recent paper, Allen, Casper, and Ottewill have found a simple
analytic formula for the $\gamma$ values of string loops in a
particular class [18].  String loops in this class have a-loops which
lie along a line, and b-loops which lie in the plane orthogonal to that
line.  In particular, when the b-loop takes the shape of an $N_b$-sided
regular polygon, $\gamma_{N_b}$ is given by
$$\gamma_{N_b}=32\bigg(1-\cos({2\pi\over N_b})\bigg)\bigg({1\over 2}
N_b\ln N_b+\sum_{j=2}^{N_b-1}j\ln(j) \cos({2\pi \over N_b}j)\bigg).
\eqno(7.7)$$
When the b-loop takes the shape of a perfect circle,
$\gamma$ is found to be $$\gamma=16\int_0^{2\pi}dx{1-\cos x\over
x}\approx 39.002454.  \eqno(7.8)$$ Since an $N_b$-sided polygon becomes
a perfect circle in the limit as $N_b$ goes to infinity, the difference
between equations (7.7) and (7.8) give $\Delta(N_b)$.  Figure 7 of
reference [18] shows a Log-Log plot of $\Delta(N_b)$.  From this plot
one finds that, in this case, $\Delta(N_b)$ falls off as $N_b^{-2}$.
The reason that the $\gamma$ values from the piecewise linear loops
converge to the smooth loop limit faster in this case than in the
previous cases is most likely because the a-loop in this case is
already piecewise linear.  While we do not know how the errors
in $\gamma_N$ scale with
increasing $N$ in the general case, it seems reasonable to conjecture
that the the errors fall off as $N^{-1}$ for large $N$.

As a point of interest, it takes only 14 seconds to calculate $\gamma$
for a loop with $N=N_a+N_b=32$ on a Sun-4 workstation (SS2).  The
calculation time for $\gamma$ scales roughly as $N^{4}$.  The
speed of this alogrithm makes it feasible to calculate $\gamma$ for
loops with large numbers of segments $N_a$ and $N_b$.  It is also
possible to rapidly calculate $\gamma$ for very large numbers of loops
with moderate values of $N_a$ and $N_b$.

In this section we have shown that in all cases where previously
published numerical methods have given reliable results for $\gamma$,
these results are in good agreement with those given by our exact
formulae.  The large number of both piecewise linear and smooth loop
trajectories for which our formulae have confirmed previously published
results gives us confidence that our method is correct.  In cases where
our method yields result that disagree with previously published
results, we have shown that our results are correct.  We have shown
this by identifying the errors in the previously published work and by
showing that our results are consistent with those given by other
independent methods such as the FFT method of Allen and Shellard [14]
and/or a corrected implementation of the numerical method used by the
original author(s).  Since our results have been correct for every
trajectory tested so far, we have confidence that our formulae provide
a reliable method for calculating the power radiated in gravitational
waves for arbitrary cosmic string loops.

\vskip 0.3in
\centerline{SECTION 8: CATALOG OF LOOPS}
\vskip 0.05in
This section gives a short catalog of piecewise linear loop
trajectories and their $\gamma$ values.  This catalog is intended to
give a number of simple cases which might prove useful in testing
future analytic or numerical methods.  The a- and b-loops which define
these trajectories are regular polygons formed with small numbers of
segments $N_a$ and $N_b$.  With each pair of a- and b-loops we form a
two parameter family of loop trajectories.  The two parameters ($\phi$
and $\theta$) describe the relative orientation of the a- and b-loop.
There is nothing special about these loops other than that they are
simple piecewise linear trajectories.

The first set of trajectories we consider are defined by a- and b-loops
consisting of 2 and 3 segments respectively.  The three segment b-loop
has the shape of an equilateral triangle.  (The simplest case, where
the a- and b-loops each have just two segments is discussed in the
previous section.)  The a-loop is taken to lie along the z-axis.  One
kink on the a-loop is positioned at the origin; the parameter $u=0$ at
this kink.  The other kink (at $u=1/2$) is positioned above the first
kink and has coordinates $(0,0,1/2)$.  For the b-loop, we again
position one kink at the origin and set the parameter $v=0$ at that
kink.  The position of the other two kinks depends on the parameters
$\phi$ and $\theta$.  When $\phi=\theta=0$, the b-loop lies in the x-z
plane.  The kink at $v=1/3$ has coordinates $(-1/6,0,\sqrt{3}/6)$ and
the kink at $v=2/3$ has coordinates $(1/6,0,\sqrt{3}/6)$.  When $\phi$
and $\theta$ are not zero, the position of the b-loop is found as
follows.  First, the b-loop is rotated by the angle $\phi >0$ about the
z-axis (counter-clockwise when viewed from large positive $z$).  After
the $\phi$ rotation, the b-loop is then rotated by the angle $\theta
>0$ about the x-axis (counter-clockwise when viewed from large positive
$x$).  Values of $\gamma$ for the trajectories defined by these a- and
b-loops are given in Table 1 for several values of the angles $\phi$
and $\theta$.

\medskip\hfil
\table{\bf Table 1}
$\theta\setminus\phi$ ! $0^{\circ}$  ! $18^{\circ}$ ! $36^{\circ}$
! $54^{\circ}$ ! $72^{\circ}$ \rr
$18^{\circ}$ 	      ! 59.33        ! 59.80        ! 61.21
! 63.51        ! 66.30        \rr
$36^{\circ}$  	      ! 53.81        ! 54.56        ! 56.86
! 60.93        ! 67.45        \rr
$54^{\circ}$ 	      ! 49.35        ! 50.15        ! 52.56
! 56.40        ! 60.72        \rr
$72^{\circ}$  	      ! 46.70        ! 47.54        ! 50.12
! 54.47        ! 60.36        \rr
$90^{\circ}$          ! 45.83        ! 46.70        ! 49.37
! 54.12        ! 62.25
\caption{}
\medskip
\noindent
In generating Table 1, we have been careful to avoid certain values of
$\phi$ and $\theta$ for which the a- and b-loops have special relative
positions.  In particular, if the a- and b-loops are exactly co-planer,
the operator $D$ becomes singular.  In this case, accurate values of
$\gamma$ may still be found by examining trajectories where the angles
$\phi$ and $\theta$ deviate very slightly from their desired values.
However, since the $\gamma$ values given in this section are meant to
be ``benchmark" values for future work, we have not included such cases
here.  In the following tables, certain pairs of angles are omitted for
similar reasons.

The second set of trajectories we consider are defined by a- and
b-loops consisting of 3 segments each.  Both the a- and b-loops are
equilateral triangles.  The position of the b-loop depends on the two
parameters $\phi$ and $\theta$ in exactly the same way as the b-loop in
the first set of trajectories.  The a-loop is placed in the same
position the b-loop has for parameter values $\phi=\theta=0$.  Values
of $\gamma$ for the trajectories defined by these a- and b-loops are given
in Table 2 for several values of the angles $\phi$ and $\theta$.

\medskip\hfil
\table{\bf Table 2}
$\theta\setminus\phi$ ! $18^{\circ}$ ! $36^{\circ}$ ! $54^{\circ}$
! $72^{\circ}$ ! $90^{\circ}$ \rr
$18^{\circ}$ 	      ! 100.85       ! 90.65        ! 74.48
! 64.80        ! 58.92        \rr
$36^{\circ}$  	      ! 82.00        ! 76.01        ! 70.97
! 64.82        ! 59.98        \rr
$54^{\circ}$ 	      ! 72.61        ! 66.51        ! 63.68
! 61.54        ! 59.95        \rr
$72^{\circ}$  	      ! 67.04        ! 61.24        ! 59.12
! 58.80        ! 60.02        \rr
$90^{\circ}$          ! 63.97        ! 58.53        ! 56.80
! 57.12        ! 59.75
\caption{}
\medskip

The third set of trajectories we consider are defined by a- and b-loops
consisting of 2 and 5 segments respectively.  The two segment a-loop is
identical to the a-loop used in the first set of trajectories.  The
b-loop is taken to be a pentagon.  One kink on the pentagon is
positioned at the origin and is chosen to have parameter value $v=0$.
When $\phi=\theta=0$, the b-loop lies in the x-z plane, and is
positioned so that the kink at $v=1/5$ has coordinates $(-{1\over
5}\cos(\pi/5),0,{1\over 5}\sin(\pi/5))$, the kink at $v=2/5$ has
coordinates $({1\over 5}(\sin(\pi/10)-\cos(\pi/5)),0,{1\over
5}(\cos(\pi/10)+\sin(\pi/5)))$, and so on.  When $\phi$ and $\theta$
are not equal to zero, the b-loop is rotated in exactly the same manner
as for the previous two sets of loop trajectories.   Values of $\gamma$
for the trajectories defined by these a- and b-loops are given in Table
3 for several values of the angles $\phi$ and $\theta$.

\medskip\hfil
\table{\bf Table 3}
$\theta\setminus\phi$ ! $0^{\circ}$  ! $18^{\circ}$ ! $36^{\circ}$
! $54^{\circ}$ ! $72^{\circ}$ \rr
$18^{\circ}$ 	      ! 63.52        ! 64.15        ! 66.04
! 69.28        ! 74.52        \rr
$36^{\circ}$  	      ! 54.07        ! 54.99        ! 57.78
! 62.31        ! 67.63        \rr
$54^{\circ}$ 	      ! 47.69        ! 48.74        ! 52.02
! 57.78        ! 66.36        \rr
$72^{\circ}$  	      ! 44.20        ! 45.30        ! 48.74
! 54.98        ! 64.04        \rr
$90^{\circ}$          ! 43.10        ! 44.20        ! 47.69
! 54.08        ! 63.95
\caption{}
\medskip

The fourth set of trajectories we consider are defined by a- and
b-loops consisting of 5 and 3 segments respectively.  The a-loop is a
pentagon placed in the same position as the b-loop in the third
trajectory set for parameter values $\phi=\theta=0$.  The b-loop is an
equilateral triangle whose position is given in terms of the parameters
$\phi$ and $\theta$ in exactly the same way as the b-loops used in the
first and second trajectory sets.  Values of $\gamma$ for the trajectories
defined by these a- and b-loops are given in Table 4 for several values
of the angles $\phi$ and $\theta$.

\medskip\hfil
\table{\bf Table 4}
$\theta\setminus\phi$ ! $18^{\circ}$ ! $36^{\circ}$ ! $54^{\circ}$
! $72^{\circ}$ \rr
$18^{\circ}$ 	      ! 84.69        ! 75.43        ! 67.82
! 62.44        \rr
$36^{\circ}$  	      ! 77.37        ! 71.03        ! 65.71
! 61.86        \rr
$54^{\circ}$ 	      ! 70.11        ! 65.69        ! 62.72
! 60.42        \rr
$72^{\circ}$  	      ! 64.41        ! 61.37        ! 61.13
! 61.33        \rr
$90^{\circ}$          ! 60.49        ! 58.21        ! 60.49
! -----
\caption{}
\medskip

The final set of trajectories we consider are defined by a- and b-loops
consisting of 5 segments each.  Both loops are taken to be pentagons.
The a-loop is identical to the a-loop used in the fourth set of
trajectories.  The b-loop is in the same position as the a-loop when
$\phi=\theta=0$.  When $\phi$ and $\theta$ are not zero, the b-loop is
rotated in the same manner as in the previous sets of trajectories.
Values of $\gamma$ for the trajectories defined by these a- and b-loops
are given in Table 5 for several values of the angles $\phi$ and
$\theta$.

\medskip\hfil
\table{\bf Table 5}
$\theta\setminus\phi$ ! $18^{\circ}$ ! $36^{\circ}$ ! $54^{\circ}$
! $72^{\circ}$ \rr
$18^{\circ}$          ! 114.46       ! 94.04        ! 80.22
! 68.84        \rr
$36^{\circ}$  	      ! 93.52        ! 82.49        ! 72.06
! 65.40        \rr
$54^{\circ}$ 	      ! 77.15        ! 72.94        ! 67.22
! 62.74        \rr
$72^{\circ}$  	      ! 67.11        ! 64.76        ! 64.05
! 62.98        \rr
$90^{\circ}$          ! 61.24        ! 59.47        ! 61.24
! -----
\caption{}
\medskip

The five sets of loop trajectories along with the $\gamma$ values given
in this section are intended as ``benchmark values" for future analytic
or numerical work.

\noindent
\vskip 0.3in
\centerline{CONCLUSION}
\vskip 0.05in
We have derived a new method for calculating the power emitted in
gravitational radiation by cosmic string loops.  This method yields an
{\it exact analytic formula} in the case of piecewise linear cosmic
string loops.  By increasing the number of segments used, piecewise
linear string loops can approximate any cosmic string loop {\it
arbitrarily closely}.  Our formula (derived in sections 5 and 6)
involves nothing more complicated than log and arctangent functions.
No numerical integrations are required.  Further, since our formula is
exact, there is no need to estimate any contribution to $\gamma$ from
the ``tail" of an infinite sum.  The error introduced when approximating
smooth loop trajectories by piecewise linear trajectories has been
investigated.  It is found that this error typically falls off
as $N^{-1}$ for large $N$, although in at least some cases it falls
off faster, as $N^{-2}$.  We believe that for ``generic" loops the
error scales as $N^{-1}$.  Using a computer to evaluate the
approximately $N^4$ terms in our formula, we can determine values of
$\gamma$ more accurately and more efficiently than by previously
published methods.

We have tested the results of our formula against all previously
published radiation rates for different loop trajectories.  Section 7
contains a detailed comparison of the results given by our new method
to those reported by previous authors.  In every case, our formula is
found to give the correct result.  In many cases our results are in
good agreement with the published data.  However there are also a
number of cases where our results do not agree with those previously
published.  There are, in fact, a number of cases where conflicting
results have been published for the same trajectories.  In the cases
where a disagreement was found, we have identified the errors made in
the published work which led to the incorrect results.  In most cases,
the errors in the published values of $\gamma$ are a result of
underestimating the contribution of the tail of the infinite sum in
(7.3).  The incorrect values of $\gamma$ which have been published are
typically 25 to 50 percent below the correct results.  We are confident
that our formula gives the correct results because, in every case, we
have shown our results to be consistent with those given by independent
methods.

We intend to use this exact formula in future work, for example to
repeat some of the work of Scherrer, Quashnock, Spergel and Press [12]
concerning the distribution of $\gamma$ values of non-self-intersecting
loops.  In addition, we plan to show how this formula may be
modified to yield similar analytic results for the linear
momentum radiated by cosmic string loops [19].

\noindent
\vskip 0.3in
\centerline{ACKNOWLEDGMENTS}
\vskip 0.05in
This work was supported in part by NSF Grant No. PHY-91-05935 and a
NATO Collaborative Research Grant.  The work of P.C. was also supported
in part by Department of Education Grant No. 144-BH22.  We would like
to thank C. Burden, R. Durrer, and T. Vachaspati for useful
correspondence concerning their work.  We are particularly grateful to
Adrian Ottewill for a number of useful conversations concerning this
work, including the tests in (6.5) and (6.6), the definition of the
operator $D$, and for providing an independent test of the formulae.

\vfil
\break
\noindent
\vskip 0.2in
\centerline{FIGURE CAPTIONS}
\vskip 0.2in

Figure 1.  For piecewise linear loops, the a- and b-loops
consist of straight segments.  The segments are joined
together at kinks where $\vec a'(u)$ and $\vec b'(v)$ are
discontinuous.  The kinks on the a-loop are labeled by the
index $i$.  The spatial position of kink $i$ is $\vec a_i$.
The value of the coordinate $u$ at the $ith$
kink is $u_i$.  The segments on the a-loop are also labeled
by the index $i$; the $ith$ segment being the one between
the $ith$ and the $(i+1)th$ kink.

\vskip 0.2in
\noindent
Figure 2.  The delta function, $\delta(f(x,y,z))$, which appears
in (5.2) only has support when $f(x,y,z)=0$.  Solving
$f(x,y,z)=0$ for $z(x,y)$ we find that the surface $z(x,y)$
consists of a pair of disconnected hyperbolic sheets.  The hyperbolic
sheets are separated by the plane $\beta_1 x+\beta_2 y+\beta_4=0$,
where the denominator of (6.4) vanishes.  The
intersection of a $z=$constant plane with these sheets will
be a hyperbola in the x-y plane.  The integration volume for
(5.2) is a box with opposite corners $(0,0,0)$ and $(\Delta x,
\Delta y,\Delta z)$.

\vskip 0.2in
\noindent
Figure 3.  The integrals (6.9) in the expression for the radiated
power are over the rectangular region bounded by the corners (0,0)
and $(\Delta x, \Delta y)$.
Each integral contains three step functions.  In the region of
integration, the first two step functions are both non-zero only in
the shaded regions between the ``Top" and ``Bottom" hyperbolae.
The third step function is on in the region which includes the ``future"
branches if $s=1$ or the ``past" branches if $s=-1$.
The vertical asymptote of the Bottom (Top)
hyperbola is shown as a dashed line that lies at $x=\phi_B$ ($x=\phi_T$).

\vskip 0.2in
\noindent
Figure 4.   The five different fundamental types of integrals that
arise when evaluating (6.9).  The type depends upon the relative
positions of the two hyperbolic curves $y(x,0)$ (labeled B for
``Bottom") and $y(x,\Delta z)$ (labeled T for ``Top") and the
rectangular region bounded by the opposite corners $(0,0)$ and $(\Delta
x,\Delta y)$.  The five possibilities are shown using the future
branches of the hyperbolae (the case for $s=+1$).  Each type of
integral may also occur with the past branches of the hyperbolae (for
$s=-1$).

\vskip 0.2in
\noindent
Figure 5.  The solid line is a plot of the analytic formula (7.1) for
$\gamma (\phi)$ for piecewise linear trajectories in which the a- and
b-loops are composed of just two segments each.  The dots show values
of $\gamma$ given by our formulae for a range of angles $\phi$.  The
agreement between the two results is so close that the dots had to be
enlarged in the figure to distinguish them from the solid line plot of
(7.1).

\vskip 0.2in
\noindent
Figure 6.  The solid curves show numerical values of $\gamma$ for
piecewise linear loop approximations of $M=1$, $L=3,5$ Burden loops.
The piecewise linear a-loops were constructed to have $N_a=16L$
segments.  The piecewise linear b-loops each had $N_b=16M=16$
segments.  The open circles show the published results of Burden [10],
the crosses show the published results of Quashnock and Spergel [13],
and the triangles show the results of the FFT method of Allen and
Shellard [14].  We find excellent agreement among all four sets of
results.

\vskip 0.2in
\noindent
Figure 7.  Numerical values of $\gamma$ for some $L=M=1$ Burden loops.
The solid line shows the results of our new method.  The open circles
show the results of Burden, the crosses show the results of Vachaspati
and Vilenkin, and the triangles show the results of the FFT method.
There is excellent agreement among these four sets of results.
Durrer's results are shown as open diamonds.

\vskip 0.2in
\noindent
Figure 8.  The $\gamma$ values reported by Durrer (open diamonds)
compared to the sum of the first 50 terms (small crosses) in (7.3) for
several $L=M=1$ Burden loops.  Including an estimate of the
contribution to $\gamma$ from the infinite tail of the sum results in
significantly larger values of $\gamma$ (large crosses).  The solid
line shows the values of $\gamma$ found by our formulae.

\vskip 0.2in
\noindent
Figure 9.  Values of $\gamma$ for piecewise linear approximations to
the $L=M=1$ Burden loops using different numbers of segments ($N_a$,
$N_b$).  The number of segments ($N_a$, $N_b$) used for the curves A,
B, C, D are respectively, (6,5), (11,10), (16,15) and (36,35).  The
Burden loops are accurately approximated over the range $\phi\in [10
,160]$ degrees when $N_a\ge 16$ and $N_b\ge 15$.  Regions $\phi <10$
and $\phi >160$ where $\gamma$ begins to diverge require a larger
number of segments before the approximation becomes accurate.

\vskip 0.2in
\noindent
Figure 10.  Relative errors in $\gamma$ for the ($N_a$, $N_b$)=(6,5),
(11,10), (16,15) and (26,25) piecewise linear loop approximations of
the $L=M=1$ Burden loops with respect to the ($N_a$, $N_b$)=(36,35)
approximation.  The relative error $\epsilon$ of each set of loops is
calculated by summing $|(\gamma^\alpha_\phi-
\gamma^{\rm{E}}_\phi)/(\gamma^\alpha_\phi+\gamma^{\rm{E}}_\phi)|$ over
the values $\phi=5,10,\dots,175$ degrees and then dividing by the
number of terms in the sum.  Here, $\alpha$ denotes which set of loops
are being compared (i.e. ($N_a$, $N_b$)=(6,5), (11,10), etc.) and E
denotes the ($N_a$, $N_b$)=(36,35) loops.  Most of the error is due to
loops with values of $\phi$ near 0 or 180 degrees.  Increasing the
number of segments from ($N_a$, $N_b$)=(16,15) to (36,35) causes the
average value of $\gamma$ to change by less than 3 percent.

\vskip 0.2in
\noindent
Figure 11.  Values of $\gamma$ are shown for the loop trajectories
given in (7.4) with $\alpha=0.5$.  The results given by our formulae
are shown by the solid line.  The results of the FFT method are shown
as triangles.  There is good agreement between the FFT method and our
new method.  Durrer's results are shown as dots while the results of
Vachaspati and Vilenkin are shown as crosses.  These last two sets of
results are inaccurate because the rate of convergence of the sum (7.3)
was estimated incorrectly.  The open circles show the sum of the first
300 terms of (7.3) as given by the FFT method.  These circles should be
taken as lower bounds on the values of $\gamma$.

\vskip 0.2in
\noindent
Figure 12.  The power spectrum for the trajectory (7.4) with
$\alpha=0.5$ and $\phi=\pi/2$ found using the FFT method.  The $P_n$
are shown in units of $G\mu^2$.  The slope of the curve for large $n$
shows that the sum (7.3) falls off as the power law $P_n\propto
n^{-1.25}$ for large $n$.  The logarithm is to base 10.

\vskip 0.2in
\noindent
Figure 13.  Values of $\gamma$ are compared for the loop trajectories
given in (7.4) with $\alpha=0.01$.  The results of our formulae are
shown by the solid line.  The results of the traditional numerical
method are shown as crosses.  The two methods are in good agreement.
The values of $\gamma$ when $\alpha=0$ are shown by the dashed line.

\vskip 0.2in
\noindent
Figure 14.  Values of $\gamma$ given by our formulae for the
trajectories (7.5) for three sets of parameters $(p,q)$ and a range of
angles $\phi$.  Curves A, B and C give the results for trajectories
with $(p,q)=$(0.6,0.4), (0.4,0.8) and (0.9,0.9) respectively.  These
results are consistent with the claim by Garfinkle and Vachaspati that
the trajectories (7.5) have $\gamma$ values on the order of 100.  The
results of the FFT method with $(p,q)=$(0.9,0.9) are shown by the
triangles, and should be compared to curve "C".  There is good
agreement between the two methods.

\vskip 0.2in
\noindent
Figure 15.  The $\gamma$ values from a series of increasingly accurate
piecewise linear loop approximations to two ($L=M=1$) Burden loops.
The total number of linear segments in each approximation is given
by $N$.  Both sets of $\gamma$ values quickly converge to their
asymptotic limits.

\vskip 0.2in
\noindent
Figure 16.  The function $\Delta(N)$ for two ($L=M=1$) Burden loops.
Each point shows the difference between the $\gamma$ value for
a piecewise linear approximation with $N$ segments and the numerical
estimate of $\gamma$ in the $N=\infty$ limit.  Both of the solid lines
have a slope of -1, showing that the errors fall off as $N^{-1}$.  The
logarithm is to base 10.

\vfil
\eject
\centerline{REFERENCES}
\item{1.} T. W. B. Kibble, J. Phys. A 9, 1387 (1976); T. W. B. Kibble,
G. Lazarides, and Q. Shafi, Phys. Rev. D 26, 435 (1982).
\item{2.} Y. B. Zel'dovich, Mon. Not. R. Astron. Soc. 192, 663 (1980).
\item{3.} A. Vilenkin, Phys. Rev. D 24, 2082 (1981); A. Vilenkin, Phys.
Rep. 121, 263 (1985).
\item{4.} A. Albrecht and N. Turok, Phys. Rev. Lett. 54, 1868 (1985);
Phys. Rev. D 40, 973 (1989); A. Albrecht, in {\it The Formation and
Evolution of Cosmic Strings}, Proceedings of the Symposium, Cambridge,
England, 1989, edited by G. W. Gibbons, S. W. Hawking, and T. Vachaspati
(Cambridge University Press, Cambridge, England, 1989).
\item{5.} D. Bennett and F. Bouchet, Phys. Rev. Lett. 60, 257 (1988);
63, 2776 (1989); Astrophys. J. 354, L41 (1990); in {\it The Formation and
Evolution of Cosmic Strings} [4]; Phys. Rev. D 41, 2408 (1990).
\item{6.} B. Allen and E. P. S. Shellard, Phys. Rev. Lett. 64, 119 (1990);
E. P. S. Shellard and B. Allen in {\it The Formation and Evolution of Cosmic
Strings} [4].
\item{7.} R. R. Caldwell and B. Allen, Phys. Rev. D 45, 3447 (1992).
\item{8.} R. Durrer, Nucl. Phys. B328, 238 (1989).
\item{9.} T. Vachaspati and A. Vilenkin, Phys. Rev. D 31, 1052 (1985).
\item{10.} C. Burden, Phys. Lett. 164B, 277 (1985).
\item{11.} D. Garfinkle and T. Vachaspati, Phys. Rev. D 36, 2229 (1987).
\item{12.} R. J. Scherrer, J. M. Quashnock, D. N. Spergel, and W. H. Press,
Phys. Rev. D 42, 1908 (1990)
\item{13.} J. M. Quashnock and D. N. Spergel, Phys. Rev. D 42, 2505 (1990).
\item{14.} B. Allen and E. P. S. Shellard Phys. Rev. D 45, 1898 (1992).
\item{15.} S. Weinberg, {\it Gravitation and Cosmology} (Wiley, New York,
1972), equation 10.4.13.
\item{16.} J. D. Jackson, {\it Classical Electrodynamics} (Wiley, New York,
1975), equation 12.133.
\item{17.} T. Vachaspati (private communication).
\item{18.} B. Allen, P. Casper, and A. Ottewill, {\it Analytic Results for
the Gravitational Radiation from a Class of Cosmic String Loops},
preprint \# WISC-MILW-94-TH-13, To appear in Phys. Rev. D 1994.
\item{19.} B. Allen, P. Casper, and A. Ottewill, {\it Closed Form Expression
for the Momentum Radiated from Cosmic String Loops},
preprint \# WISC-MILW-94-TH-15, unpublished.

\bye \end